\documentclass[11pt, a4paper]{article}
\usepackage{fullpage}
\usepackage{amsfonts}
\usepackage{amssymb}
\usepackage{amsmath}
\usepackage{amsthm}
\usepackage{graphicx}
\usepackage{bm}
\usepackage[makeroom]{cancel}
\usepackage{enumitem}
\usepackage{url}
\usepackage[margin=.9in]{geometry}
\usepackage{amsopn}
\usepackage{mathtools}
\usepackage{hyperref}
\usepackage{doi}
\usepackage{cite}
\usepackage{overpic}
\usepackage[symbol]{footmisc}
\usepackage{empheq}

\usepackage{listings}
\usepackage{xcolor}
\definecolor{codegreen}{rgb}{0,0.6,0}
\definecolor{codegray}{rgb}{0.5,0.5,0.5}
\definecolor{codepurple}{rgb}{0.58,0,0.82}
\definecolor{backcolour}{rgb}{0.95,0.95,0.92}
\lstdefinestyle{mystyle}{
    backgroundcolor=\color{backcolour},   
    commentstyle=\color{codegreen},
    keywordstyle=\color{magenta},
    numberstyle=\tiny\color{codegray},
    stringstyle=\color{codepurple},
    basicstyle=\ttfamily\normalsize,
    breakatwhitespace=false,         
    breaklines=true,                 
    captionpos=b,                    
    keepspaces=true,                 
    numbers=left,                    
    numbersep=3pt,                  
    showspaces=false,                
    showstringspaces=false,
    showtabs=false,                  
    tabsize=4,
}
\renewcommand{\vec}[1]{\mathbf{#1}}

\newcommand{\norma}[1]{\left|\left|{#1}\right|\right|}

\newcommand{\dpart}[2]{\frac{\partial {#1}}{\partial {#2}}}
\newcommand{\der}[2]{\frac{d {#1}}{d {#2}}}

\begin{document}
\begin{center}
    \Large \bf Multi-Fidelity Learning with Shallow Recurrent Decoders for Multi-Physics Applications
\end{center}
\begin{center}
    Stefano Riva$^{1,2}$,
    Carolina Introini$^{2}$,
    J. Nathan Kutz$^{1}$,
    Antonio Cammi$^{3,2}$
\end{center}
\begin{center}
    \scriptsize{
    ${}^1$ Autodesk Research, London, UK \\  
    ${}^2$ Department of Energy, Nuclear Engineering Division, Politecnico di Milano, Milan, Italy \\  
    ${}^3$ Department of Mechanical and Nuclear Engineering and Emirates Nuclear Technology Center, Khalifa University, Abu Dhabi (127788), United Arab Emirates}
\end{center}

\begin{abstract}
    In reactor physics, neutronics and, more broadly, coupled multi-physics phenomena can be modelled at different fidelity levels, according to the needs of the user or the tolerance levels required in practice. High-fidelity models based on the Boltzmann transport equation, multi-group diffusion, or coupled computational fluid dynamics are computationally demanding, whereas simplified models, such as point kinetics or zero-dimensional lumped formulations, can be evaluated efficiently at the cost of neglecting spatial detail. The computational intractability of detailed models translates into a scarcity of high-fidelity data and an abundance of low-fidelity data, motivating the development of multi-fidelity (MF) learning strategies able to map between the two. This work extends Shallow Recurrent Decoders (SHRED), a machine learning architecture that reconstructs high-dimensional fields from sparse or global time-series measurements, to multi-fidelity reduced-order modelling, in which the input trajectories are provided by a low-fidelity model rather than physical sensors. The resulting MF-SHRED architecture is assessed on three benchmark problems of progressively increasing fidelity gap: (i) a two-group point-kinetics-to-diffusion mapping on the LRA benchmark geometry, in which the lumped model is projection-based; (ii) a non-linear reaction-advection-diffusion system of six chemical species, in which the low-fidelity lumped model arises from a global balance; and (iii) the coupled neutronics-thermal-hydraulics multi-physics model of the Molten Salt Fast Reactor (MSFR), in which a lumped, delay-differential model of the circulating fuel is mapped to a 22-field, three-dimensional solution. Across all three cases, MF-SHRED reconstructs the high-fidelity fields with relative errors generally below a few percent, closely approaching the truncation error of the underlying proper orthogonal decomposition basis, while reducing the computational cost by two to three orders of magnitude relative to the corresponding high-fidelity solver. Notably, MF-SHRED performs comparably to, and in some cases better than, the original sparse-sensor SHRED formulation, and is shown to correct biases of the low-fidelity driving model rather than merely reproducing them at higher resolution. These results support the use of MF-SHRED as a general-purpose, non-intrusive reduced-order modelling strategy for reactor physics and multi-physics applications, particularly for design, control, and safety analysis tasks that must be carried out before a facility is instrumented or even built.
\end{abstract}

\section{Introduction}\label{sec: intro}

The accurate prediction of neutronics behaviour in nuclear reactors is fundamental for safety analysis, design optimization, and monitoring applications. However, the numerical solution of high-fidelity mathematical models, represented by systems of coupled partial differential equations such as the one arising from the discretisation of the Boltzmann transport equation \cite{DuderstadtHamilton}, is computationally demanding or, for multi-query scenarios, potentially intractable even on high-performance computers. This drawback limits their use for multi-query scenarios, as in optimisation tasks or uncertainty quantification \cite{rozza_model_2020}, in which multiple evaluations of the model are required. This computational cost is also a major concern in multi-physics (MP) settings, where neutronics must be coupled with thermal-hydraulics at the least (as well as with chemistry, thermo-mechanics, fluid-structure interface) \cite{demaziere_6_2020}. The most egregious example of the intrinsic multi-physics nature of nuclear systems is perhaps the Molten Salt Fast Reactor (MSFR), a circulating-fuel reactor in which neutronics, thermo-fluid-dynamics and precursor drift all interact together in a strict multi-physics sense \cite{serp_molten_2014, aufiero2014development}.

In the nuclear field, this computational shortcoming persists despite the advancements in computational hardware and software: Reduced-Order Modelling (ROM) techniques \cite{rozza_model_2020, brunton_data-driven_2022, quarteroni2015reduced} have been extensively studied by the nuclear engineering community to provide fast yet reliable approximations of high-fidelity solutions \cite{GERMAN2022104148}. Research started from projection-based methods \cite{benner2015survey}, in which the known governing equations are projected onto a low-dimensional subspace, before shifting towards non-intrusive methods, in which the reduced model is learnt directly from the available data \cite{Rozzareview}. Data-driven methods offer greater flexibility compared to projection-based methods: being independent from the starting equations, the same framework can be applied to different problems with only a few modifications on the implementation. This greater flexibility comes at the cost that, if not properly trained, data-driven models, and in particular neural networks-based ones, may yield unphysical solutions, especially when deployed in new parameter regimes or in extrapolation mode \cite{SUN2026118604}.

Alongside ROM methodologies, reactor physics widely adopts low-fidelity models, obtained by applying approximations and simplifying assumptions to the starting high-fidelity PDE system. Diffusion theory from transport is the chief example \cite{DuderstadtHamilton}, or Reynolds-Averaged Navier Stokes (RANS) from Direct Numerical Simulations in Computational Fluid Dynamics (CFD) \cite{versteeg2007introduction}, and lumped models based on the Point Kinetics (PK) equations have been extensively used for control; plant simulators based on lumped or one-dimensional approaches (including system codes such as RELAP5) are also widely used for system-level analysis and stability assessments \cite{LORUSSO2023112627}. In a sense, the PK equations can be considered a projection-based ROM of the neutron transport equation following Henry' approach, that is, by projecting the equations using the adjoint operator \cite{Henry01011958, VALOCCHI2020107702}; multi-physics coupling can then be obtained by using balance equations averaged in space and algebraic relations to describe the effect of the various physics on the system reactivity \cite{DuderstadtHamilton, demaziere_6_2020}.

The advancements in computational power have led the nuclear industry to adopt more and more accurate models, including multi-group diffusion, transport and computational fluid-dynamics models. Despite these advancements, such models are still not feasible for multi-query analysis or online applications, including parameter optimization, uncertainty quantification,  control and monitoring. As such, high-fidelity datasets remain scarce (and often unavailable to the public), whereas there is an abundance of (also historical) low-fidelity data. This dichotomy makes nuclear reactor physics an interesting field for studying multi-fidelity (MF) learning approaches. Briefly, the key idea of MF is to use information coming from multiple sources of different accuracy and computational cost. The recent advances in Machine Learning (ML) and data-driven ROM offer new perspectives for combining low- and high-fidelity descriptions of reactor physics phenomena \cite{conti2024_mf, TORZONI2023110376, conti2025progressivemultifidelitylearningphysical}. 

An ever-increasing number of ROM and ML techniques, with varying results, have been applied to problems of the nuclear industry. This work follows the ongoing research of the authors on the SHallow REcurrent Decoders (SHRED) paradigm  \cite{williams2022data, shredrom, RIVA2025105928}, which has been proved to be capable of estimating a high-dimensional field from sparse time-series measurements with great accuracy, for different scientific fields, from temperature forecast to plasma dynamics to bio-mechanics to nuclear physics \cite{riva2025_parametricMSFR, riva2026modelsexperimentsshallowrecurrent, Faraji_2025, ebers2023leveraging, gao2026uqshreduncertaintyquantificationshallow, tomasetto2026realtimeoptimalcontrolshallow, pomarico2026shallowrecurrentdecoderdynamic, verso2026surrogatemodelsnuclearfusion}. Compared to other state estimation methods, SHRED presents several promising advantages: (i) it can work with very few measurements; (ii) it is agnostic to sensor positions; (iii) it requires minimal hyperparameter tuning. In its original formulation, SHRED uses sparse sensor measurements to reconstruct the full state of the system, making this framework an optimal solution for the monitoring tasks within the development of digital twins \cite{NURETH25_evol}.

However, design and optimisation tasks are performed before building the actual facility, hence measurements are not available, making the state estimation process for control and monitoring not directly possible. Given this apparent limitation, Faraji et al. \cite{Faraji_2025} have shown that it is possible to train a SHRED model also using global measurements for the prediction of plasma dynamics; therefore, it is legitimate to investigate the possibility of generating a reduced-order model of some high-fidelity model, such as diffusion or transport, starting from temporal trajectories coming from a low-fidelity model, such the PK or a lumped model. While this idea has been shown to be feasible when the fidelity gap between the low- and high-fidelity models is relatively small (for example, when both models describe the same physics at different spatial resolutions, as in the PK-to-diffusion mapping for reactor kinetics), it remains an open question whether SHRED can bridge substantially larger fidelity gaps. Such gaps arise, for instance, when the low-fidelity model lacks fields present in the high-fidelity solution or attempts to approximate inherently spatial phenomena, such as in the lumped model of the MSFR that carries no information regarding the spatially-resolved velocity or precursor motion. Addressing this question is mandatory to establish the robustness of multi-fidelity learning using SHRED (MF-SHRED) as a general-purpose reduced-order modelling tool, to obtain accurate estimations of high-fidelity models with the computational cost of low-fidelity ones, thus deriving a physics-based surrogate model suitable for reactor physics applications before deployment.

This work investigates this possibility, at first considering a benchmark test case based on the LRA benchmark geometry from the Argonne Benchmark Book \cite{argonne_book}, with the goal of mapping a PK model to a two-group diffusion solution, both linear. The second test case is a non-linear reaction-diffusion-transport problem of multiple species, in which the low- and high-fidelity models are separated both in physical content and computational cost. Finally, the third test case involves the mapping between a lumped, zero-dimensional model and a high-fidelity, spatial MP solution for the Molten Salt Fast Reactor. The main contributions of this work can be summarised as follows: i) the extension of the SHRED architecture to multi-fidelity reduced-order modelling, mapping low-fidelity lumped models to high-fidelity spatially-resolved solutions; ii) the assessment of MF-SHRED across three benchmark problems characterised by progressively large fidelity gaps; iii) a quantification of the trade-off between accuracy and computational acceleration achieved by MF-SHRED compared to the corresponding high-fidelity solvers; iv) evidence that global, spatially-averaged low-fidelity inputs can provide high-fidelity state reconstruction even when they do not contain the same physical fields as the target high-fidelity model.

Multi-fidelity techniques have recently gained traction in the nuclear engineering community as a way to combine the accuracy of high-fidelity multi-physics solvers with the speed of simplified or lower-resolution models. For thermal-hydraulic transient analysis, hybrid reduced-order surrogates combining regression and neural-network closures have been used to accelerate system-level codes such as SPACE \cite{BANG2025111002}. In the context of fuel assembly thermal-hydraulics, a cross-scale multi-fidelity ROM combining proper orthogonal decomposition with transfer learning has been proposed to reconstruct large-scale flow-field data for fuel rod bundles from small-scale training data only \cite{Kang}. Closer to the multi-physics setting considered in this work, reduced-order modelling of Molten Salt Reactor thermal-fluid dynamics has been pursued combining parametric POD with higher-order dynamic mode decomposition \cite{en19102387}, building on the multi-physics ROM literature for circulating-fuel systems discussed above \cite{GERMAN2022104148}. Multi-fidelity Kriging has also been applied to accelerate digital-twin-oriented material property prediction for accident-tolerant fuel concepts, combining physics-based low-fidelity models with sparse high-fidelity experimental data \cite{Kobayashi2021}. At the system level, multi-fidelity uncertainty quantification frameworks have been developed for problems in which the low- and high-fidelity models are parametrised differently, with application to finite-element models of spent nuclear fuel \cite{N2024116546}, while multi-fidelity reinforcement learning has been explored for shape and power-distribution optimisation in nuclear micro-reactors. These works confirm the growing relevance of multi-fidelity strategies for nuclear applications, though, to the authors' knowledge, none of them addresses the specific problem tackled in this paper: mapping a lumped, lower-dimensional dynamical model directly to a full spatially-resolved field using a sensor-free, sequence-to-state learning architecture.

Beyond the nuclear domain, multi-fidelity learning has been the subject of extensive research in the broader scientific machine learning community, dating back to the foundational survey by Peherstorfer, Willcox, and Gunzburger \cite{Gunzburger}, which systematised multi-fidelity methods for uncertainty propagation, inference, and optimisation. More recently, the combination of dimensionality reduction with multi-fidelity regression has become a common strategy for approximating high-dimensional PDE solutions from cheaper, lower-fidelity data: Conti et al. \cite{conti2024_mf, conti2025progressivemultifidelitylearningphysical}, already discussed above, integrate POD with multi-fidelity long short-term memory networks, while operator-learning approaches such as DeepONet have been extended to multi-fidelity residual-learning settings for reduced-order modelling \cite{Demo} and to data-efficient prediction of unsteady flows via physics-guided subsampling \cite{YANG2025118254}. Multi-fidelity surrogates have also been developed for structural health monitoring, combining model order reduction with artificial neural networks to infer damage indicators from sparse, noisy measurements \cite{TORZONI2023110376}, and Bayesian neural network approaches have been proposed to propagate uncertainty across fidelity levels in aerodynamic load prediction \cite{Kerleguer_2024}. A recent review specifically surveys the intersection of machine-learning-based reduced-order modelling and multi-fidelity field reconstruction in computational fluid dynamics, covering both data-driven mapping strategies and super-resolution-inspired approaches \cite{WU2026123156}.

Multi-fidelity techniques have recently gained traction in the nuclear engineering community as a way to combine the accuracy of high-fidelity multi-physics solvers with the speed of simplified or lower-resolution models. For thermal-hydraulic transient analysis, hybrid reduced-order surrogates combining regression and neural-network closures have been used to accelerate system-level codes such as SPACE \cite{BANG2025111002}. In the context of fuel assembly thermal-hydraulics, a cross-scale multi-fidelity ROM combining proper orthogonal decomposition with transfer learning has been proposed to reconstruct large-scale flow-field data for fuel rod bundles from small-scale training data only \cite{Kang}. Closer to the multi-physics setting considered in this work, reduced-order modelling of Molten Salt Reactor thermal-fluid dynamics has been pursued combining parametric POD with higher-order dynamic mode decomposition \cite{en19102387}, building on the multi-physics ROM literature for circulating-fuel systems discussed above \cite{GERMAN2022104148}. Multi-fidelity Kriging has also been applied to accelerate digital-twin-oriented material property prediction for accident-tolerant fuel concepts, combining physics-based low-fidelity models with sparse high-fidelity experimental data \cite{Kobayashi2021}. At the system level, multi-fidelity uncertainty quantification frameworks have been developed for problems in which the low- and high-fidelity models are parametrised differently, with application to finite-element models of spent nuclear fuel \cite{N2024116546}, while multi-fidelity reinforcement learning has been explored for shape and power-distribution optimisation in nuclear micro-reactors. These works confirm the growing relevance of multi-fidelity strategies for nuclear applications, though, to the authors' knowledge, none of them addresses the specific problem tackled in this paper: mapping a lumped, lower-dimensional dynamical model directly to a full spatially-resolved field using a sensor-free, sequence-to-state learning architecture.

The paper is organised as follows. Section \ref{sec: shred} presents the SHRED architecture and its extension to multi-fidelity reduced-order modelling: Section II.A introduces the parametric SHRED and the multi-fidelity extension , while Section II.B provides the theoretical justification for treating a low-fidelity model as a global measurement of the corresponding high-fidelity field. Section \ref{sec: num-res} reports the numerical results on three benchmark problems of progressively increasing fidelity gap. Finally, Section \ref{sec: concl} summarises the main conclusions of this work and outlines directions for future developments, whereas Appendix \ref{app: params-testcases} provides the detailed physics models and parameters for all three test cases.

\section{Shallow Recurrent Decoders for Multi-Fidelity Learning}\label{sec: shred}

The framework represented by Shallow Recurrent Decoders has been shown to be a powerful technique that can map sparse measurements to the corresponding high-dimensional state; it has been successfully applied to different physics, ranging from climate modelling, fluid dynamics, electrical networks, nuclear fission and fusion reactor modelling, and biology \cite{williams2022data, shredrom, RIVA2025105928, Rude2025, verso2026surrogatemodelsnuclearfusion, verso2026applicationparametricshallowrecurrent, pomarico2026shallowrecurrentdecoderdynamic}. The SHRED architecture is composed of a recurrent unit, typically a Long Short-Term Memory network (LSTM), and a shallow decoder (SDN); the former learns a suitable latent space from temporal trajectories, whereas the latter maps this learnt space to the high-dimensional one. Both components of SHRED have two layers each (with 64 neurons each for the recurrent unit and 350 and 400 neurons for the latter), making the overall network relatively small\footnote{This same architecture has been successfully applied to most of the aforementioned applications, without the need of changing the number of neurons and layers.}. Previous studies have also shown that the network can work in compressive mode \cite{Faraji_2025, shredrom}, in which the output of the shallow decoder lies in the reduced space obtained by Singular Value Decomposition (SVD): the high-dimensional state can then be retrieved by projecting the learnt latent dynamics onto the spatial modes computed with the SVD. This solution reduces both the training database and the output size, and consequently the training time, allowing for training even on personal computers.

Beyond spatial-temporal datasets, the SHRED architecture has been extended to parametric scenarios \cite{shredrom, riva2025_parametricMSFR} for the state $u(\vec{x}; t, \boldsymbol{\mu})$, where $\boldsymbol{\mu}\in\mathcal{D}\subset\mathbb{R}^{N_\mu}$ collects operating or input conditions, such as the reactivity, the Reynolds number or the position of the control rods in a reactor. This extension requires minimal modifications to the standard formulation: the time-delay embedding required by the LSTM naturally accommodates multiple trajectories associated with different parameter values, and the vector $\boldsymbol{\mu}$ can be included as additional input when known a priori, or estimated as output when inferred from measurements \cite{shredrom}. The critical step lies in the compressive training phase. For a fixed parameter $\boldsymbol{\mu}_p$, let $\mathbb{X}^{\boldsymbol{\mu}_p}\in\mathbb{R}^{\mathcal{N}_h\times N_t}$ denote the snapshot matrix associated with $\boldsymbol{\mu}_p$, with $\mathcal{N}_h$ the spatial degrees of freedom and $N_t$ the number of saved time steps. The SVD provides a rank-$r$ basis $\mathbb{U}^{\boldsymbol{\mu}_p}$ and the reduced coefficients $\mathbb{V}^{\boldsymbol{\mu}_p}=\left(\mathbb{U}^{\boldsymbol{\mu}_p}\right)^T\mathbb{X}^{\boldsymbol{\mu}_p}\in\mathbb{R}^{r\times N_t}$, which encode the temporal dynamics. For a parametric dataset composed of $P$ realization of parameters, a common reduced basis spanning the entire solution manifold must be constructed. When the dimension of the problem is sufficiently small \cite{riva2025_parametricMSFR}, this can be achieved by stacking the snapshots
\begin{equation}
    \mathbb{X} = \left[\mathbb{X}^{\boldsymbol{\mu}_1}|\mathbb{X}^{\boldsymbol{\mu}_2}| \dots | \mathbb{X}^{\boldsymbol{\mu}_{P}}\right]\in\mathbb{R}^{\mathcal{N}_h\times N_t\cdot P},
    \label{eqn: parametric-stacked-snapshots}
\end{equation}
and performing the SVD on $\mathbb{X}$. For larger datasets, hierarchical or incremental variants of the SVD on the non-stacked matrices can be adopted instead \cite{iwen_distributed_2016, halko_finding_2010}. Once the global basis $\mathbb{U}$ is available, the compressive training proceeds as in the non-parametric case: for each parameter $\boldsymbol{\mu}_p\in\mathcal{D}$, the input measurements are lagged as
\begin{equation}
    \left[\vec{y}^{\boldsymbol{\mu}_p}_{k}, \dots , \vec{y}^{\boldsymbol{\mu}_p}_{k-L}\right], \qquad k=1,\dots,N_t,
    \label{eqn: parametric-lagging}
\end{equation}
and the parameter sets $\Xi_{\text{train}}^{\boldsymbol{\mu}}$, $\Xi_{\text{valid}}^{\boldsymbol{\mu}}$ and $\Xi_{\text{test}}^{\boldsymbol{\mu}}$ are defined by splitting $\boldsymbol{\mu}$ into training, validation and test subsets. Denoting by $\vec{f}_R$ and $\vec{f}_D$ the recurrent and decoder maps, the parametric compressive loss function reads
\begin{equation}
    \widetilde{\mathcal{J}} = \sum_{(\boldsymbol{\mu}_i, k) \in \Xi_{\text{train}}^{\boldsymbol{\mu}} \times \Xi_{\text{train}}^{t}} \norma{\mathbb{U}^T\vec{u}(\boldsymbol{\mu}_i, t_k) - \vec{f}_D\left(\vec{f}_R\left(\vec{y}^{\boldsymbol{\mu}_i}_{k}, \dots , \vec{y}^{\boldsymbol{\mu}_i}_{k-L}\right)\right)}_2^2.
    \label{eqn: pshred-loss}
\end{equation}
At inference, the trained network provides a reduced-order surrogate valid for unseen parameter instances within the training range, enabling both parametric interpolation and, with adequate coverage of $\mathcal{D}$, moderate extrapolation. This capability has been demonstrated for accidental transients in circulating-fuel reactors \cite{riva2025_parametricMSFR} and motivates the present extension to multi-fidelity learning, in which the input trajectories are provided by a low-fidelity model rather than sparse sensors.

Most of the applications of SHRED in the nuclear reactor field have been for monitoring purposes in the broad framework of state estimation problems, focusing on the reconstruction from sparse measurements. This task has been successfully applied for reconstruction, predictions, and forecasting scenarios, and for multi-physics problems considering only a single observable field \cite{RIVA2025105928, NURETH25_evol, riva2025_parametricMSFR, verso2026applicationparametricshallowrecurrent}; in particular, it has been shown that very few input sensors (around 3 to 5) are enough to obtain an accurate state estimation, due to the fact that temporal histories are given as input to the recurrent unit: this ensures a deeper knowledge as stated by the Takens time-delay embedding theorem. The SHRED architecture has been implemented in Python using the \href{https://pytorch.org/}{Pytorch} package; the original code \cite{williams2022data} has been adapted for the present application, and it is openly available at \href{https://github.com/ERMETE-Lab/NuSHRED}{github.com/ERMETE-Lab/NuSHRED}, as well as the notebooks used to generate the results.

\begin{figure}[tp]
    \centering
    \includegraphics[width=1\linewidth]{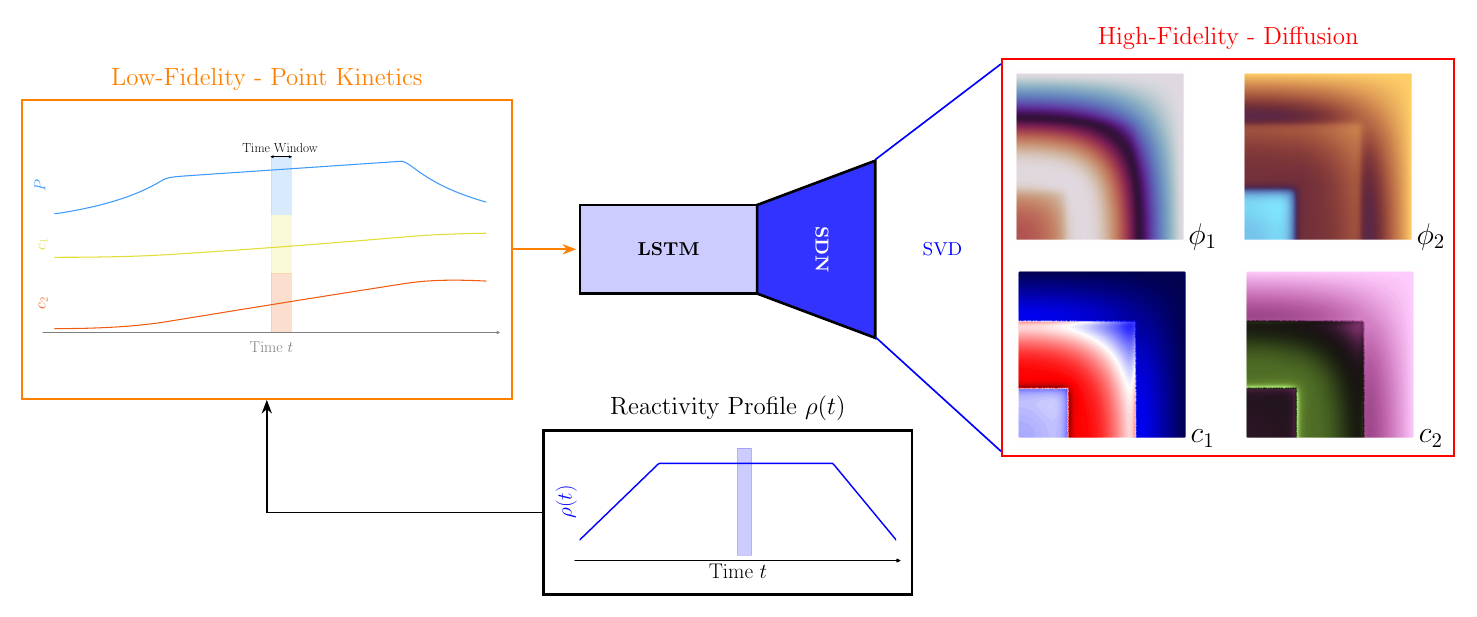}
    \caption{Scheme of the Multi-Fidelity SHRED (MF-SHRED), where the solution of the point kinetics equations is mapped to the high-fidelity diffusion model.}
    \label{fig: shred-mf}
\end{figure}

\subsection{Extension to Multi-Fidelity}\label{sec: shred-mf-ext}

This work extends the parametric SHRED framework to multi-fidelity reduced-order modelling \cite{conti2024_mf, conti2025progressivemultifidelitylearningphysical}. Let $\mathcal{M}_{\text{HF}}$ and $\mathcal{M}_{\text{LF}}$ denote a high-fidelity and a low-fidelity model, respectively, providing solutions $\vec{u}_{\text{HF}}(t, \boldsymbol{\mu})\in\mathbb{R}^{\mathcal{N}_h}$ and a vector of lumped or global observables $\vec{u}_{\text{LF}}^{\boldsymbol{\mu}}(t)\in\mathbb{R}^{s}$, with $s\ll \mathcal{N}_h$. Given the HF snapshots, the parametric SVD basis $\mathbb{U}$ and the reduced coefficients $\vec{v}_{\text{HF}}^{\boldsymbol{\mu}}(t_k)=\mathbb{U}^T\vec{u}_{\text{HF}}(\boldsymbol{\mu}, t_k)$ are computed as in Eqs.~\eqref{eqn: parametric-stacked-snapshots}--\eqref{eqn: pshred-loss}; the Multi-Fidelity SHRED (MF-SHRED) network is then trained to map the lagged LF trajectories to the HF reduced dynamics, i.e.
\begin{equation}
    \vec{v}_{\text{HF}}^{\boldsymbol{\mu}}(t_k) \approx \vec{f}_D\left(\vec{f}_R\left(\vec{u}^{\boldsymbol{\mu}}_{\text{LF},k}, \dots , \vec{u}^{\boldsymbol{\mu}}_{\text{LF},k-L}\right)\right),
    \label{eqn: mf-shred-map}
\end{equation}
and the HF state is recovered as $\hat{\vec{u}}_{\text{HF}}(\boldsymbol{\mu}, t_k)=\mathbb{U}\,\vec{f}_D\left(\vec{f}_R\left(\vec{u}^{\boldsymbol{\mu}}_{\text{LF},k}, \dots , \vec{u}^{\boldsymbol{\mu}}_{\text{LF},k-L}\right)\right)$. Even though local measurements have been generally used as input for SHRED, global observables can be adopted within the same formalism \cite{Faraji_2025}, since the recurrent unit only requires time-series data and does not need information on how they are collected. Accordingly, the LF solution of $\mathcal{M}_{\text{LF}}$ can replace sparse sensors as network input, enabling SHRED to act as a non-intrusive ROM of $\mathcal{M}_{\text{HF}}$ for multi-query applications: the spatial behaviour is encoded in the SVD modes, whereas the temporal and parametric dynamics are learnt by the neural network, from the similar (for an accurate LF model) trajectory of the LF model. This idea is depicted in Figure~\ref{fig: shred-mf}, where the Point Kinetics solution is mapped to the multi-group diffusion model; once trained, a new reactivity profile $\boldsymbol{\mu}^\star$ is obtained by integrating $\mathcal{M}_{\text{LF}}$ and decoding with MF-SHRED, rather than solving $\mathcal{M}_{\text{HF}}$. In principle, this procedure can be embedded to multi-step fidelity learning, where multiple corrections of the accuracy can be included.

\subsection{Low-Fidelity models as global measurements}\label{sec: mf-shred-explained}

The measurement procedure can be mathematically represented as a functional $\mathcal{G}$ taking as input a function $u(\vec{x}, t, \boldsymbol{\mu})$ and returning a scalar value $y(t, \boldsymbol{\mu})$ \cite{argaud_sensor_2018, riva2024multiphysics}, specifically:
\begin{equation}
    y(t, \boldsymbol{\mu}) = \mathcal{G}[u] = \int_\Omega g(\vec{x}) \, u(\vec{x}, t, \boldsymbol{\mu}) \, d\Omega,
    \label{eqn: measurement-functional}
\end{equation}
where $g$ is a suitable integration kernel. For local sensors, $g$ is generally selected as a Gaussian kernel, such that $\mathcal{G}[1]=1$, characterised by a centre of mass and a point spread, or as a Dirac delta function to identify point measurements \cite{haik_real-time_2023, ICAPP_plus2023}. From this formulation, every local measurement which is used as input for the SHRED architecture, can be represented. Furthermore, the same formulation can be adopted to extract global measurements like spatial averages over the whole domain\footnote{This is achieved by selecting $g$ as a constant function equal to $1/|\Omega|$, where $\Omega$ is the domain of integration.}; the same logic can be used to describe sectional measurements, such as the power in a specific slice of the system.

High-fidelity models are generally derived from local conservation principles on local variables, whereas lumped models satisfy global conservation laws (mass, momentum and energy). Given $u_{HF}$ be the solution of the HF model and $u_{LF}$ be the solution of the LF model, both satisfy a proper set of conservation laws; there is generally a connection between them, in particular 
\begin{equation}
    \frac{1}{|\Omega|}\int_\Omega u_{HF}(\vec{x}, t, \boldsymbol{\mu}) \, d\Omega = u_{LF}(t, \boldsymbol{\mu}) + \delta(t, \boldsymbol{\mu})
\end{equation}
where $\delta(t, \boldsymbol{\mu})$ is a discrepancy term accounting for local phenomena, affecting the global dynamics, which cannot be described with a lumped approach. It can be noticed that the left-hand side is nothing but the spatial average of the HF state or its global measure $\bar{u}_{HF}(t, \boldsymbol{\mu})$; accordingly, the LF is nothing but an approximation\footnote{If the LF model can be analytically derived from the HF they become equal, as in the pair point-kinetics and diffusion following a adjont-based approach.}, i.e., $\bar{u}_{HF}(t, \boldsymbol{\mu})\approx u_{LF}(t, \boldsymbol{\mu})$. This heuristic reasoning justify the use of SHRED for Multi-Fidelity applications in which a cheap LF model is used as input trajectory for this decoding only strategy: in particular, it shows that the good properties of the SHRED should be maintained even when sensing is no more sparse but rather global.

\color{black}

\section{Numerical Results}\label{sec: num-res}

This paper considers three different test cases of interest in the nuclear engineering community. The first one is the most simple case, where both the LF and HF models are linear and there exist an almost exact map between the two, i.e. the point kinetics and the multi-group diffusion equations. The second problems extends the methodology to a non-linear system of coupled equations for a reaction-diffusion-advection problem. Lastly, the MF-SHRED is tested on a fully coupled reactor, the Molten Salt Fast Reactor, where the discrepancy between fidelities is significant and they are both highly non-linear.

Before entering into the details, two metrics are to be defined. The mean relative error $\varepsilon_1$ of the SVD/POD coefficients with respect to the ML prediction: given $v_{j,r}^{\boldsymbol{\mu}_n}$ the $r$-th SVD coefficient at the $j-$th time step for the $n-$th parameter $\boldsymbol{\mu}_n$, and $\hat{v}_{j,r}^{\boldsymbol{\mu}_n}$ its ML prediction, the mean relative error is defined as
\begin{equation}
    \varepsilon_1^r = \frac{1}{N_s^{test}}\sum_{n=1}^{N_s^{test}}\left( \frac{1}{N_t}\sum_{j=1}^{N_t}\frac{\left|v_{j,r}^{\boldsymbol{\mu}_n} - \hat{v}_{j,r}^{\boldsymbol{\mu}_n}\right|}{\left|v_{j,r}^{\boldsymbol{\mu}_n}\right|}\right)
\end{equation}
The second metric to introduce is the average relative error $\varepsilon_2^\psi$ of generic field $\psi$, defined as the energy norm of the residual relative to the energy of the HF model, i.e.,
\begin{equation}
    \varepsilon_2^{\psi} = \frac{1}{N_s^{test}}\sum_{n=1}^{N_s^{test}}\left( \frac{1}{N_t}\sum_{j=1}^{N_t}\frac{\norma{\boldsymbol{\Psi}_j^{\boldsymbol{\mu}_n} - \hat{\boldsymbol{\Psi}}_j^{\boldsymbol{\mu}_n}}_2}{\norma{\boldsymbol{\Psi}_j^{\boldsymbol{\mu}_n}}_2}\right)
\end{equation}
where $\boldsymbol{\Psi}_j^{\boldsymbol{\mu}_n}$ is the HF solution of the field $\psi$ at the $j-$th time step for the $n-$th parameter $\boldsymbol{\mu}_n$, and $\hat{\boldsymbol{\Psi}}_j^{\boldsymbol{\mu}_n}$ its ML prediction.

\subsection{Linear Model with Neutronics: LRA 2D Benchmark}
\begin{figure}[tp]
    \centering
    \includegraphics[width=0.5\linewidth]{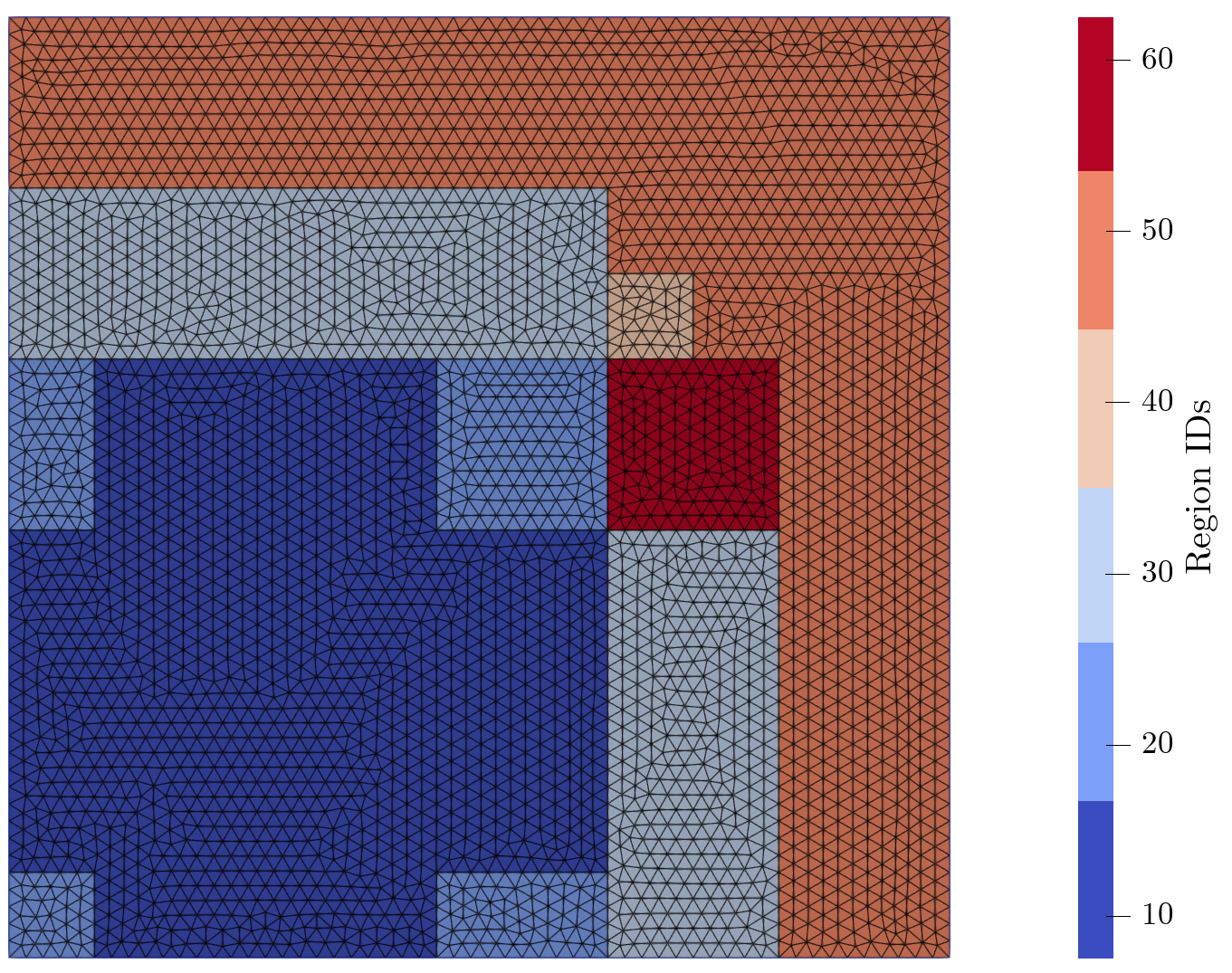}
    \caption{Geometry, numerical mesh and regions of the LRA benchmark \cite{argonne_book}. All regions except ID = 50 (dark orange) are multiplying media with different composition; transient behaviour is triggered in region ID = 60 (red, rod region) by modifying its thermal absorption cross section. }
    \label{fig: lra-domain-region}
\end{figure}

The MF-SHRED methodology has first been tested on a benchmark based on the geometry of the LRA BWR problem from the Argonne Benchmark Book \cite{argonne_book}, mapping a two-group Point Kinetics (PK) model to a two-group diffusion solution; the full formulation of both models, together with the parametric transient definitions, is reported in Appendix \ref{app: lrabench}. The geometry, shown in Figure \ref{fig: lra-domain-region} includes six regions, five of which are multiplying media (except for the external region with ID = 50, in dark orange). The group cross sections have been generated with OpenMC rather than taken from the original benchmark, to demonstrate that the approach is not tied to a specific cross-section set and can be applied to other neutronic problems with minimal changes.

Three parametric transient shapes (ramp, sinusoidal and trapezoidal) have been considered as perturbations of the  absorption cross section $\Sigma_{a,2}(\mathbf{x},t;\boldsymbol{\mu})$ of the thermal group in the rod region (ID = 60) from its nominal value, such that $\Sigma_{a,2}(\mathbf{x},t;\boldsymbol{\mu})=\Sigma_{a,2}^{(0)}(\mathbf{x})\cdot f(t;\boldsymbol{\mu})$. The interested reader is referred to Appendix \ref{app: lrabench}. The parameter $\boldsymbol{\mu}$ in $f(t;\boldsymbol{\mu})$ depends on the transient; a similar input condition is given to the PK model through reactivity $\rho(t;\boldsymbol{\mu})$. More details are provided in Appendix \ref{app: lrabench}. The dataset includes 50 parametric realisations per transient shape, up to $T_{final} = 4$ s, split into train ($\sim$72\%), validation ($\sim$8\%) and test ($\sim$20\%) sets by random sampling of the parametric space. The mesh accounts for $\mathcal{N}_h = 3770$ spatial degrees of freedom, and 4 fields are considered: the two neutron flux groups $\phi_1$ and $\phi_2$, and the two precursor groups $c_1$ and $c_2$. On the other hand, the low-fidelity input is the PK solution itself (power and two precursors groups): only three input channels are used by the recurrent unit, a deliberate choice to demonstrate that MF-SHRED does not need, in principle, a large number of sensors to perform well in this setting.

\begin{figure}[tp]
    \centering
    \includegraphics[width=1\linewidth]{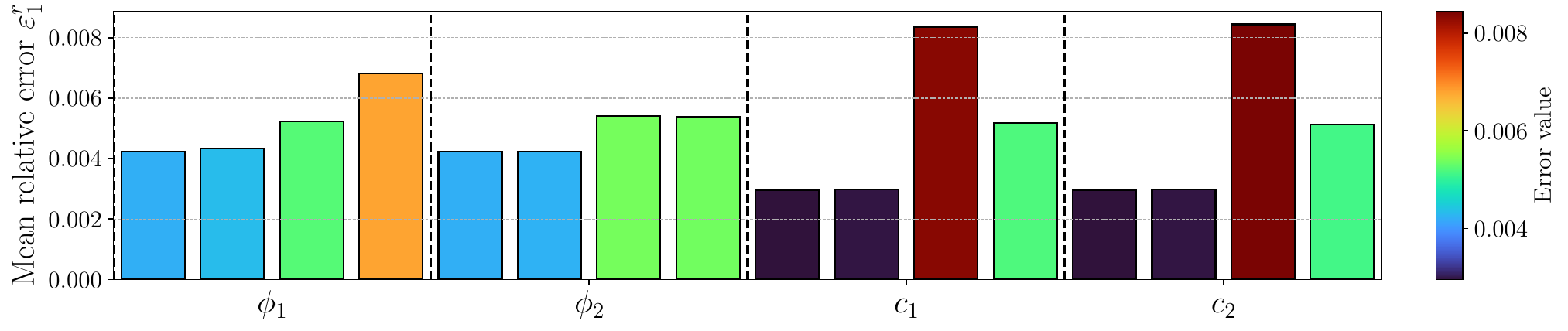}
    \caption{Mean relative error on the POD coefficients predicted by MF-SHRED for the LRA benchmark, reported per coefficient for all four fields. The lowest-order modes are reconstructed with the smallest error, while a single higher-order mode per field concentrates most of the residual, consistent with the decreasing energy content of higher-order POD modes.}
    \label{fig: podcoeff_error}
\end{figure}

The training HF snapshots are first reduced through Singular Value Decomposition, retaining only 4 modes per field (as these modes are enough to retain $99.9$\% of the information, as per decay of the singular values, see the companion Github for more information).  Training the SHRED architecture takes approximately 2 minutes on a MacBook Pro M4 using the GPU via Metal Performance Shaders (MPS), and approximately 5 minutes on CPU; following training, the MF-SHRED is tested on unknown parametric scenarios (either from ramp, sine, and trapezoid). The mean relative error on the POD latent coefficients can be computed. Figure \ref{fig: podcoeff_error} shows the mean relative error $\varepsilon_1^r$ for each coefficient of the fields of interest. For all four fields, the lowest-order modes, which are associated with the largest singular values and thus the dominant spatial patterns, are reconstructed with the smallest error (below 0.003 in most cases), while a single higher-order mode per field concentrates most of the residual error, reaching approximately 0.006 for $\phi_2$ and approximately 0.010 for both precursor groups. This behaviour confirms that the dominant, slowly-varying components of the dynamics are learnt more easily than the higher-order corrections, which carry comparatively little energy but capture faster or more localised features of the transient. Despite this mode-wise disparity, the field-level error remains low overall, confirming that the higher-order modes contribute only marginally to the reconstructed field.

Then, the predictions are compared at the high-dimensional level by projecting back the predicted POD coefficients using the SVD basis. All fields are reconstructed with error below $2\%$, specifically:
\begin{equation}
    \begin{array}{lcl}
    \varepsilon_2^{\phi_1}=0.0187\pm0.0111 & \qquad &
    \varepsilon_2^{\phi_2}=0.0134\pm0.0079\\
    \varepsilon_2^{c_1}=0.00032\pm0.00023 & \qquad &
    \varepsilon_2^{c_2}=0.00071\pm0.00051
    \end{array}
\end{equation}
It can be observed that the difference between the estimated quantities and the high-fidelity diffusion solution is overall very low, meaning that all the fields are, on average, accurately reconstructed. 

\begin{figure}[tp]
    \centering
    \includegraphics[width=1\linewidth]{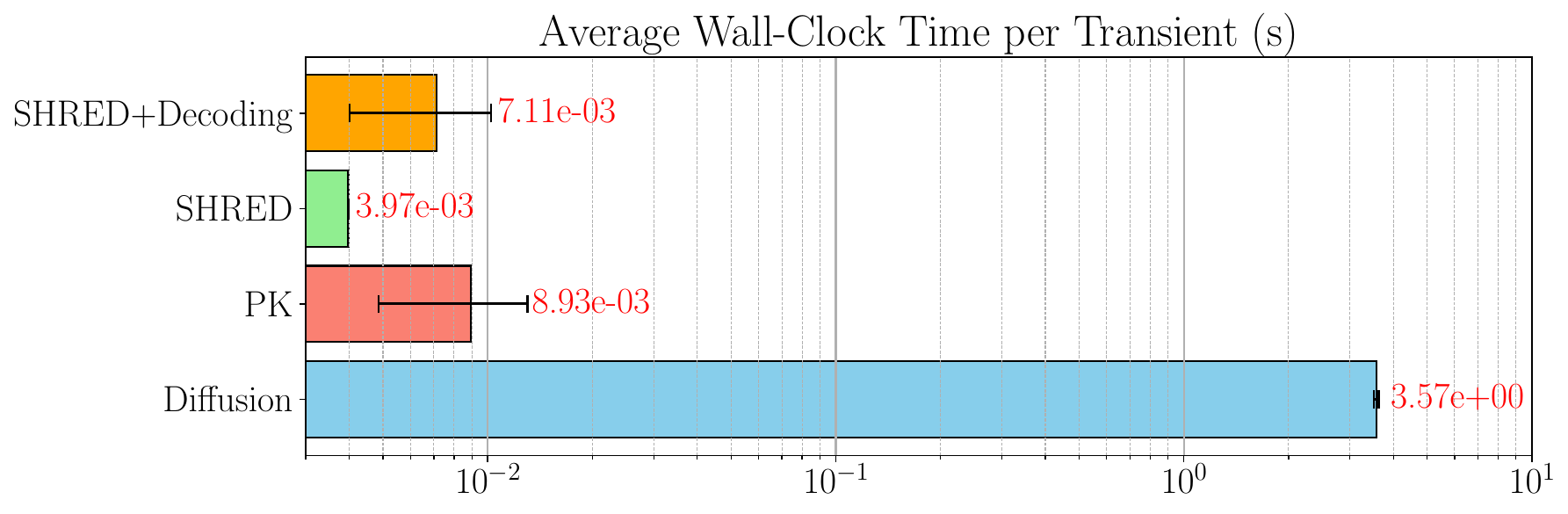}
    \caption{Average wall-clock time, in seconds, needed to obtain a solution from each modelling methods: the diffusion requires much more time, whereas the PK and the MF-SHRED (including decoding with SVD) have comparable time.}
    \label{fig: times}
\end{figure}

In addition to the accuracy of MF-SHRED, it is important to assess the efficiency of the approach, namely, its associated computational time in inferring a new HF solution. All the operations have been performed on the same machine, and it has been chosen to use the wall-clock time as a metric for the computational performance: even though the training of SHRED is typically done on laptop GPUs, being more efficient, it has been decided to use a CPU to have a fairer comparison with the finite element solution and the PK integration. Figure \ref{fig: times} compares the average wall clock time, expressed in seconds, that each model requires to obtain a full solution of the transient for a new parameter value. The diffusion finite-element solution requires approximately 3.57 s, whereas the PK integration and the MF-SHRED evaluation (including SVD decoding, a simple matrix-vector product) both take under 0.01 s, a speed-up of more than two orders of magnitude with respect to the high-fidelity solver, at comparable cost to the low-fidelity model alone.

\begin{figure}[tp]
    \centering
    \includegraphics[width=1\linewidth]{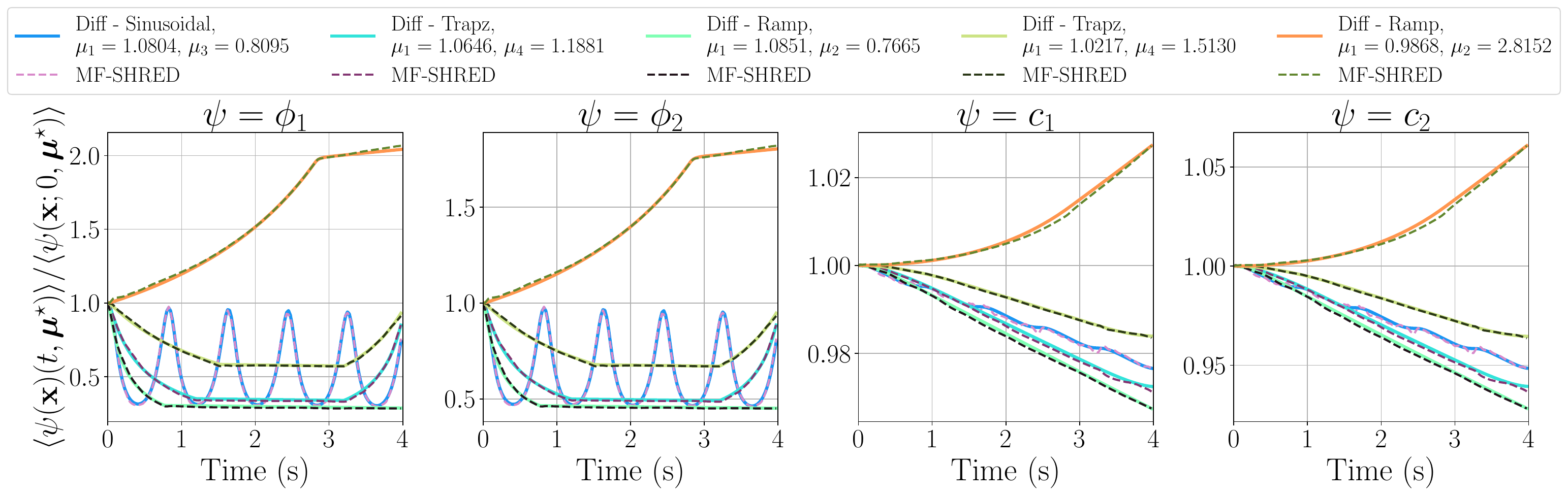}
    \caption{Temporal evolution of the normalised spatial average $\langle\psi(\mathbf{x})(t, \boldsymbol{\mu}^\star)\rangle / \langle\psi(\mathbf{x}; 0, \boldsymbol{\mu}^\star)\rangle$ of the different fields, normalised with respect to the initial condition (continuous line is the diffusion, dashed line is the MF-SHRED prediction).}
    \label{fig: fom-ave}
\end{figure}

\begin{figure}[tp]
    \centering
    \includegraphics[width=1\linewidth]{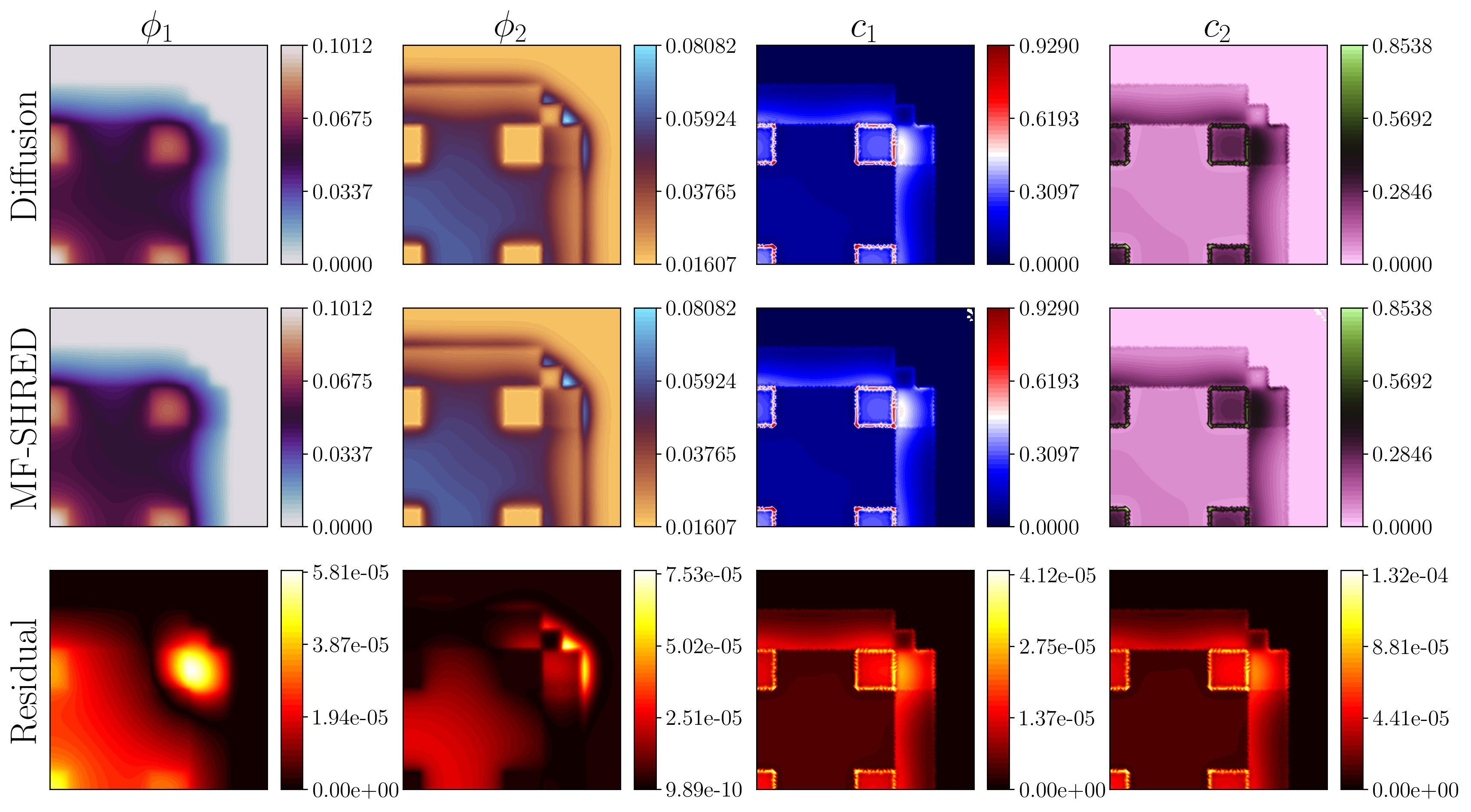}
    \caption{Contour plots (the fluxes are normalised to have total reactor power equal to 1) at the last time step for a test parameter (sinusoidal transient with $\mu_1 \approx 1.08$ and $\mu_3\approx 0.8095$) of the fast flux $\phi_1$ (first column), thermal flux $\phi_2$ (second column), first precursors group $c_1$ (third column) and second precursors group $c_2$ (fourth). From left to right: high-fidelity solution, estimation with MF-SHRED, the associated the residual field. The prediction with SHRED provides a correct local state estimation.}
    \label{fig: contour}
\end{figure}

After assessing that globally the estimation with MF-SHRED is accurate by measuring the error, it is possible to observe how the reconstruction behaves both in time and space: Figure \ref{fig: fom-ave} shows the spatial average of each estimated field over time for different parameters in the test set (from each transient type), normalised with respect to the initial condition; Figure \ref{fig: contour} shows instead the contour plot of the different fields, comparing the diffusion solution with the estimation with MF-SHRED, including the associated residual field (i.e., the absolute difference). It can be observed that the dynamics are always well predicted and that the architecture is able to discriminate the specific transient under consideration and reconstruct the state accurately \cite{PHYSOR26_shred_msfr}, without any direct knowledge of the input reactivity profile; this is not only true for the global quantities, but also locally. The decoding layers and the SVD allow for representing the spatial behaviour in a compact way, making the overall architecture generally cheap to train and thus enabling quick estimations.

Overall, these results indicate that MF-SHRED is able to reconstruct both the neutron fluxes and the precursor concentrations of the LRA benchmark with high accuracy, using only the point kinetics solution as input trajectory. The field-level errors remain below 2\% for all quantities, and the POD coefficients analysis shows that the dominant, low-order modes are essentially exact, while the residual error concentrates in a single higher-order mode per field, without significantly degrading the reconstructed field. From a computational standpoint, MF-SHRED, including SVD decoding, reproduces the diffusion solution at a cost comparable to the point kinetics model itself, more than two orders of magnitude faster than the finite element solver, while requiring only a few minutes of offline training. These results confirm that MF-SHRED can bridge a small fidelity gap, where the low- and high-fidelity models describe the same physics at different spatial resolutions. 

\subsection{Non-Linear Model with Reaction-Advection-Diffusion of Species}

The second test case considers a reaction-advection-diffusion system of six chemical species ($c_0$ to $c_5$) on a two-dimensional square domain, discretised using a finite element method, implemented using \texttt{dolfinx} \cite{BarattaEtal2023,ScroggsEtal2022,AlnaesEtal2014,BasixJoss}: the governing equations and the initial/boundary conditions for the studied case can be found in Appendix \ref{app: reactdiff}. The high-fidelity model solves the spatially-resolved PDE system, while the low-fidelity model integrates the corresponding lumped ODE system, sharing the same reaction kinetics and temporal source forcing but removing all spatial information (including the spatial transport and spatial profiles of the source terms). 

The dataset spans a Cartesian product of 10 values of the linear reaction scale $\mu_1 \in [0.5, 2]$, 5 values of the source peak time $\mu_2 \in [0.5, 3]$ and 10 values of the source pulse width $\mu_3 \in [0.2, 2]$, for a total of $N_s = 500$ parametric combinations; each of them is simulated up to $T_{final} = 10$ s, with a time step $\Delta t = 0.01$ s. The dataset is randomly split into train ($\sim$60\%), validation ($\sim$8\%) and test ($\sim$32\%) sets. Each snapshot for each field is scaled independently to $[0,1]$, for visualisation reasons. The parametric SVD basis is now built per species from the training snapshots, selecting for each specie the minimum rank $r_i$ such that the cumulative energy exceeds $99.99\%$ for each one: this results in these ranks $[7,4,7,3,4,4]$ for $c_0,\dots,c_5$ respectively, for a total latent dimension $R = \sum_i r_i = 29$.

Two SHRED variants are compared on this test case. The baseline SHRED follows the original sparse-sensor formulation, using $3$ local measurements drawn from species $c_0$, $c_3$ and $c_5$ only (to have information on the full decay chain). The MF-SHRED model instead uses 3 low-fidelity species trajectories taken from the lumped ODE solution as input. For both variants, an ensemble of $5$ independently-trained configurations is considered: for the baseline, each configuration corresponds to a different random placement of the sensors; for MF-SHRED, each configuration  corresponds to a different random subset of $3$ out of the $6$ low-fidelity species trajectories. Predictions are averaged across the ensemble to obtain the final estimation, following the same strategy used in \cite{RIVA2025105928}.
\begin{figure}[tp]
    \centering
    \includegraphics[width=1\linewidth]{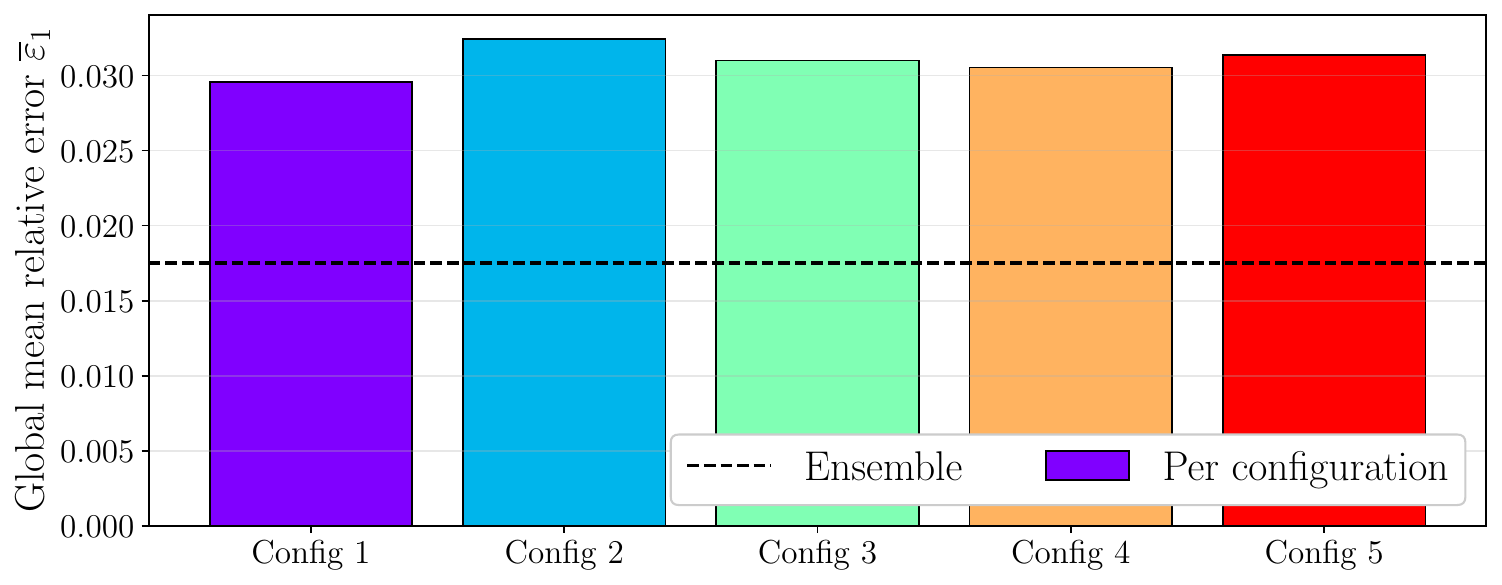}
    \caption{Mean relative error on the MF-SHRED POD coefficients for each of the 5 random low-fidelityspecies input configurations (bars), compared against the error of the ensemble-averaged prediction across all configurations (dashed line). Averaging over configurations substantially reduces the reconstruction error relative to any individual configuration.}
    \label{fig: RDA-mfshred_pod_error}
\end{figure}

Figure \ref{fig: RDA-mfshred_pod_error} shows the global mean relative error $\overline{\varepsilon}_1 = R^{-1} \sum_r \varepsilon_1^r$ on the POD coefficients obtained for the five configurations of MF-SHRED: these errors are higher than the one of the ensemble-averaged prediction, confirming that the ensemble strategy of combining multiple random low-fidelity species subsets reduces the variance of the reconstruction, at the acceptable cost of training 5 independent networks rather than 1 (acceptable given the training times of SHRED, which remains of the order of minutes even on personal, mid-end laptops). Moreover, there is very little difference in the reconstruction error across the five configurations, confirming that the MF-SHRED architecture is robust to the choice of low-fidelity species used as input, and overall agnostic also for global measurements.

\begin{figure}[tp]
    \centering
    \includegraphics[width=1\linewidth]{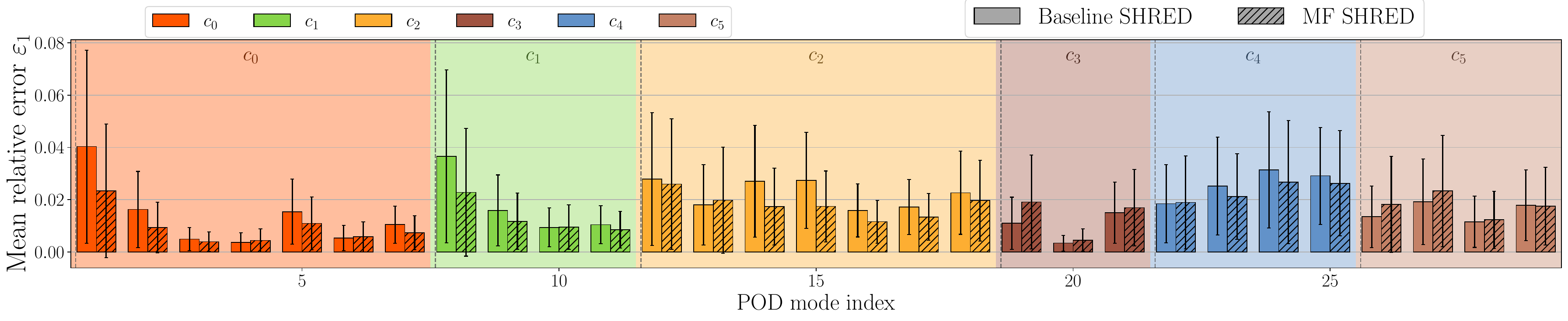}
    \caption{Mean relative error on each retained POD coefficient, comparing baseline SHRED (sparse sensors) and MF-SHRED (low fidelity species trajectories), grouped by species (separated by dashed lines). Error bars reflect the uncertainty across the ensemble of configurations.}
    \label{fig: RDA-pod_error_comparisonshreds}
\end{figure}

Figure \ref{fig: RDA-pod_error_comparisonshreds} compares instead the mean relative error on each retained POD coefficient for the baseline (lighter bars) and MF-SHRED (darker bars) models: both variants reconstruct the retained latent coefficients with comparable accuracy, as the error for the worst-reconstructed coefficient remains around $4\%$ (first coefficient for specie $c_0$).

\begin{figure}[tp] 
    \centering
    \includegraphics[width=1\linewidth]{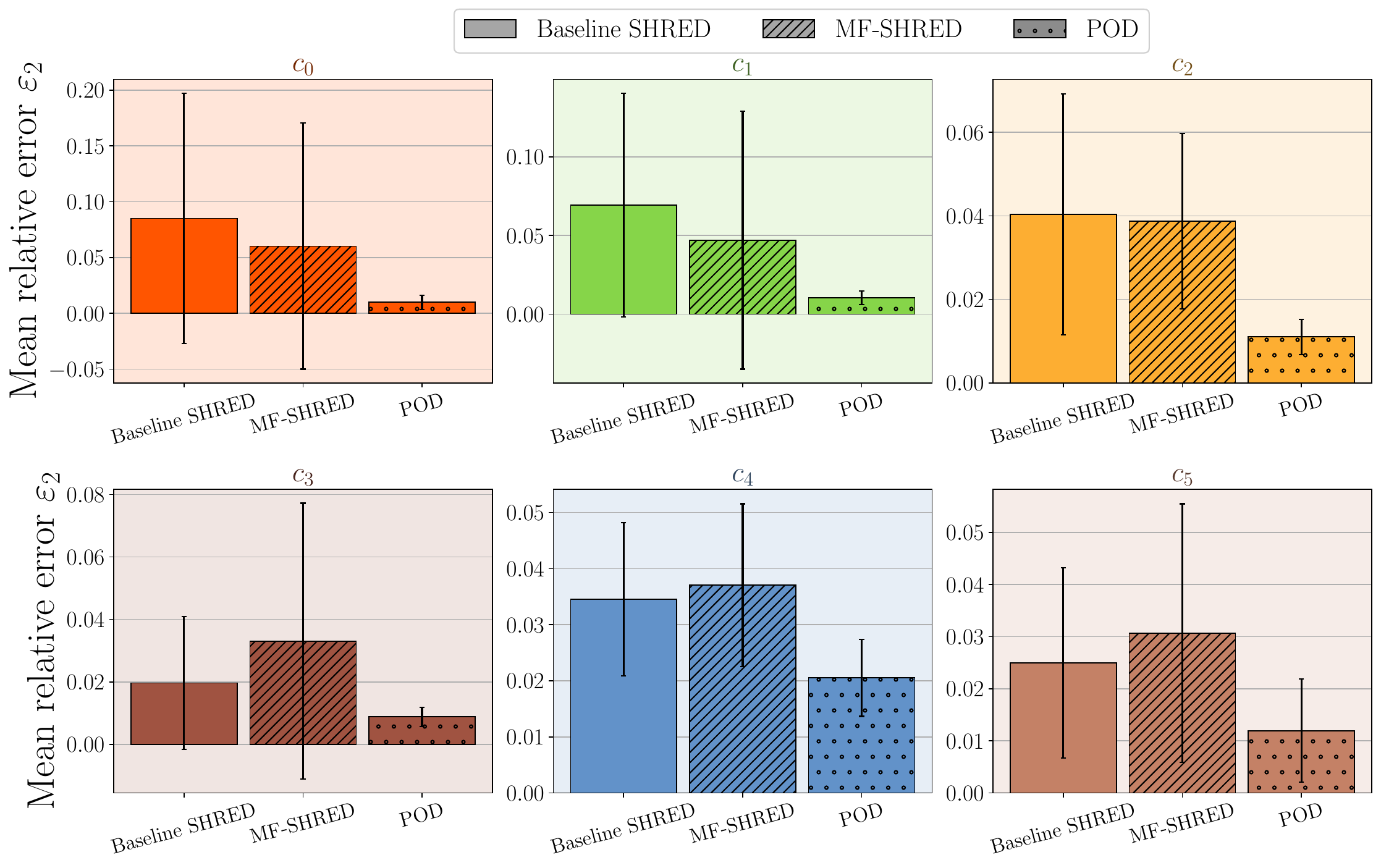}
    \caption{Mean relative reconstruction error on the high-dimensional field for each species, comparing baseline SHRED, MF-SHRED, and the POD truncation error (lower bound achievable at the retained rank). Both SHRED variants remain close to the POD floor for all species.}
    \label{fig: RDA-recon_error_comparisonshreds}
\end{figure}

\begin{figure}[tp]
    \centering
    \includegraphics[width=1\linewidth]{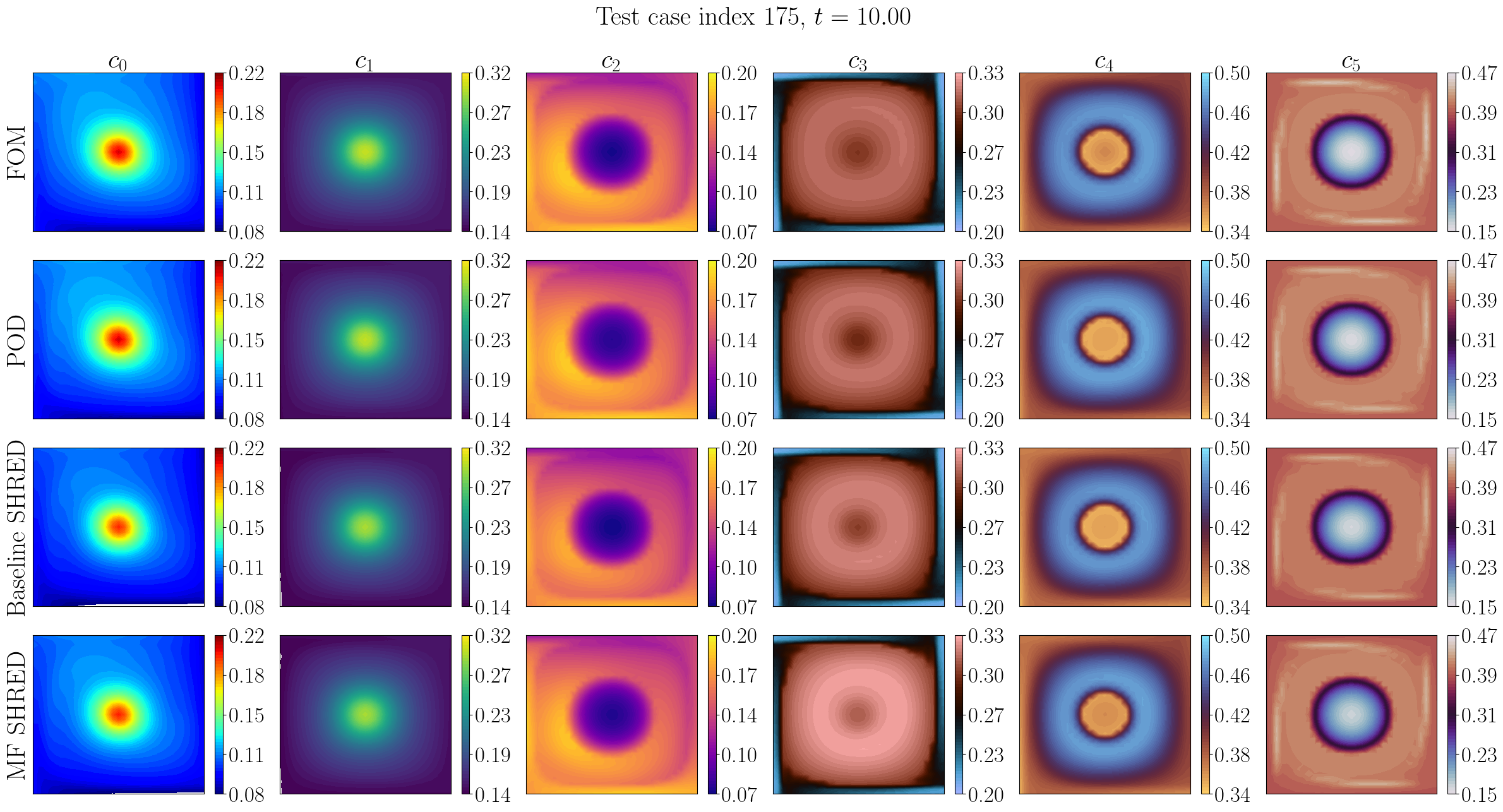}
    \caption{Spatial reconstruction of all six species for a representative test case at the final simulation time, comparing the full-order model (FOM), the POD projection, the baseline SHRED reconstruction, and the MF-SHRED reconstruction.}
    \label{fig: RDA-spatial_snapshot_comparison}
\end{figure}

Figure \ref{fig: RDA-recon_error_comparisonshreds} now compares the relative error $\varepsilon_2$ for baseline SHRED, MF-SHRED, and the POD truncation error (the best achievable error given the retained rank), once the output of SHRED has been decoded back to the starting high-dimensional space. For all six species, both SHRED variants remain close to the POD floor, with MF-SHRED performing comparably to, and in some species (such as $c_0$ and $c_1$) slightly better than, the baseline, despite having access only to spatially-averaged information rather than local point measurements. This result is consistent with the theoretical argument of Section \ref{sec: mf-shred-explained}: since the low-fidelity lumped trajectories approximate the spatial average of the high-fidelity fields, they carry comparable information content to a small set of local sensors. The qualitative agreement is confirmed in Figure \ref{fig: RDA-spatial_snapshot_comparison}, which compares the full-order, POD, baseline SHRED and MF-SHRED reconstructions for a representative test case at the final simulation time: all three reduced representations are qualitatively equal to the full-order solution for every species.

\begin{figure}[tp] 
    \centering
    \includegraphics[width=1\linewidth]{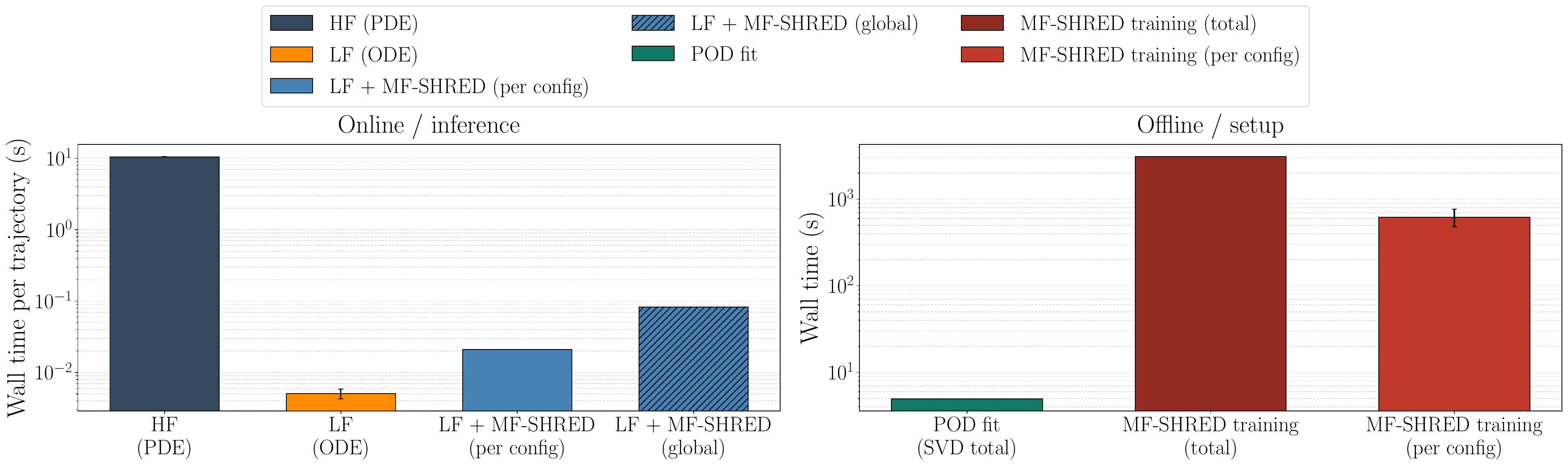}
    \caption{Computational cost comparison. Left: online wall time per trajectory for the high-fidelity solver, the low-fidelity ODE solver, and MF-SHRED, using either a single configuration or the full ensemble. Right: offline setup cost, comparing the total POD fit time against the total and per-configuration MF-SHRED training time.}
    \label{fig: RDA-computational_cost_comparison}
\end{figure}

Finally, Figure \ref{fig: RDA-computational_cost_comparison} reports the computational cost of MF-SHRED for the training (right) and deployment phase (left). Solving the high-fidelity PDE model for a parametric triplet requires $\sim 10$ seconds, whereas the lumped ODE model has basically negligible costs. More significantly, MF-SHRED remains significantly cheaper than the high-fidelity model, achieving a computational time savings of two orders of magnitude (considering all five configurations). Now considering training times, training the MF-SHRED ensemble is clearly more expensive in total than fitting the POD basis, although the mean training time per configuration is lower than the cumulative ensemble cost: regardless, this additional cost is comparable to very few high-fidelity evaluations, confirming the advantages of this approach in multi-query scenarios.
 
Overall, this test case shows that MF-SHRED remains effective even as the fidelity gap widens compared to the LRA benchmark, as now the low-fidelity model discards all spatial information entirely, retaining only six lumped concentrations. Despite this, the reconstructed high-dimensional fields remain close to the POD truncation floor for every species, and are qualitatively equal to the full-order solution. Notably, MF-SHRED achieves this accuracy while performing comparably to (and in some species marginally better than), the baseline sparse-sensor formulation, despite receiving only spatially-averaged information rather than local point measurements: the likely explanation for this behaviour is that a lumped low-fidelity trajectory acts as an approximate global measurement of the high-fidelity field. Finally, the ensemble strategy of averaging over multiple random subsets of low-fidelity species reduces the reconstruction error relative to any single configuration, at a moderate additional offline training cost. Overall, these results seems to indicate that the MF-SHRED paradigm can generalise to a case where the low- and high-fidelity models differ in dimensionality.

\subsection{Non-Linear Multi-Physics Model for the Molten Salt Fast Reactor}

The third test case considers the coupled neutronics-thermal-hydraulics multi-physics model \cite{aufiero2014development} of the Molten Salt Fast Reactor (Appendix \ref{app: msfr-gov-eqn}), mapped from a lumped, zero-dimensional low-fidelity model of the same reactor, described in Appendix \ref{app: msfr-0d}. Both models simulate the same accidental transient, an Unprotected Loss Of Flow (ULOFF), in which the primary pump coasts down exponentially with time constant $\tau \in [1, 10]$ seconds, sampled at 21 equally-spaced values (see \cite{riva2025_parametricMSFR} for more details about the dataset). The parameters are split into train, validation and test with the following ratio 71.4\%, 14.3\%, 14.3\% respectively. Compared to the previous cases, the low-fidelity model now does not share the same complete physical state as the high-fidelity solutions, as it includes only the reactor power, eight delayed-neutron precursors densities and a single lumped fuel temperature (whereas the high-fidelity model additionally resolves the velocity field, the six-group neutron flux, the pressure, the three decay-heat precursors and the turbulence quantities over the full spatial domain). Following the same parametric SVD procedure as the previous test case, a uniform rank of 10 modes is retained for each of the 22 high-fidelity fields, for a total latent dimension $R = 220$. MF-SHRED is trained on an ensemble of $10$ configurations, each using $3$ randomly-selected lumped input trajectories taken from the $10$ available lumped outputs (reactor power, eight precursor groups, fuel temperature). 

Given the wide conceptual jump in approximating a circulating fuel with a spatially-averaged values, a preliminary sanity check is performed on the lumped model, independently of the SHRED architecture: the high-fidelity fields for each parametric instance are spatially-averaged over the in-core region and compared against the corresponding low-fidelity lumped trajectories, all normalised by their initial value. 

\begin{figure}[tp]
    \centering
    \includegraphics[width=1\linewidth]{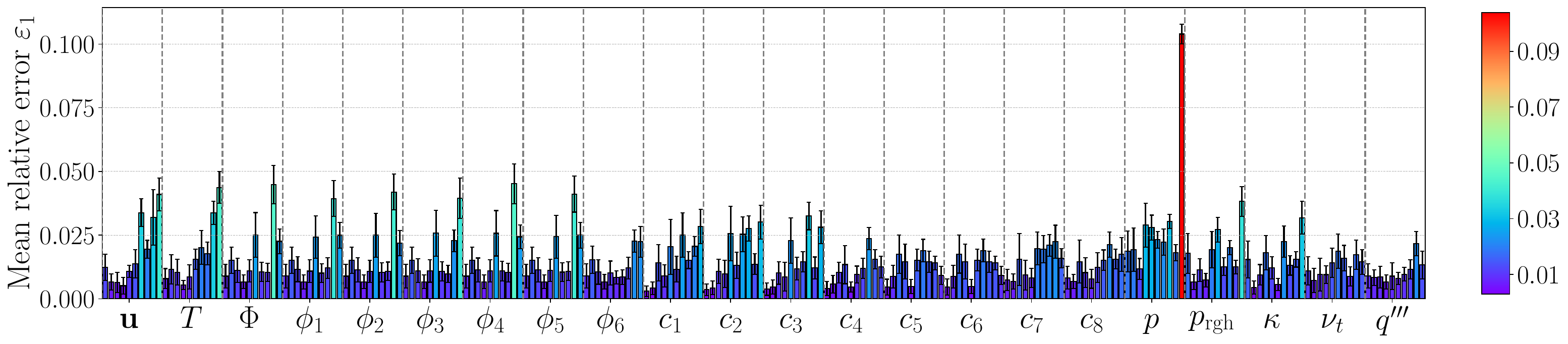}
    \caption{Mean relative error $\varepsilon_1$ on each of the 220 retained POD coefficients of the MSFR MF-SHRED model, grouped by field (vertical dashed lines) and colour-coded by the error magnitude of each mode.}
    \label{fig: MSFR-pod_coeff_error_bars}
\end{figure}

The ensemble-averaged MF-SHRED prediction achieves a mean relative error $\varepsilon_1$ of 2.12\% ($\pm$ 0.44\%) on the scaled POD coefficients across the test set. Figure \ref{fig: MSFR-pod_coeff_error_bars} shows the mean relative error for each of the 220 retained POD coefficients, grouped by field: the error remains low and fairly uniform across most fields, with a visible increase concentrated in a subset of the precursor-group modes. The highest error are recorded for higher-order coefficients, which is a desired behaviour as the lower-order coefficients contains the majority of the information of the fields, whereas higher order coefficients capture local behaviours. All errors on the coefficients remains below $5\%$, with the notable exception of the last two coefficients for pressure, which peaks at around $10\%$.

\begin{figure}[tp]
    \centering
    \includegraphics[width=1\linewidth]{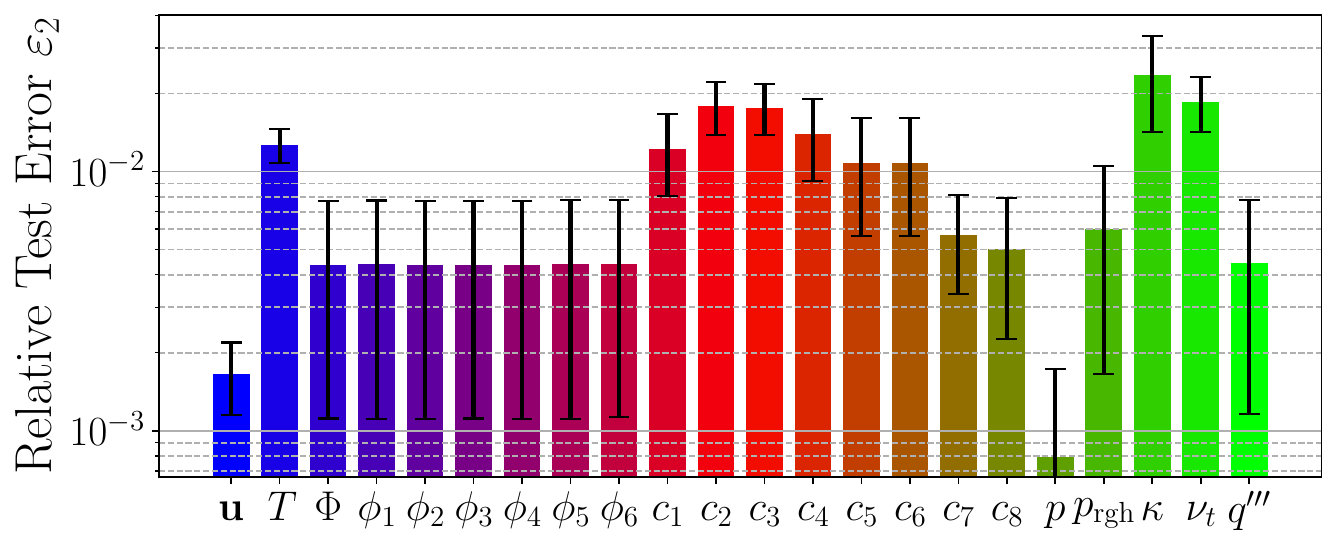}
    \caption{Mean relative reconstruction error $\varepsilon_2$ on the high-dimensional field, for each of the 22 MSFR fields, averaged over the test $\tau^\star$ values and the transient duration.}
    \label{fig: MSFR-fom_relativerrror_bars}
\end{figure}

Following the decoding step back to the high-dimensional space, Figure \ref{fig: MSFR-fom_relativerrror_bars} reports the relative reconstruction error $\varepsilon_2$ for each field, averaged over the test parameter cases and the transient duration. The velocity field and pressure are reconstructed with the lowest error, confirming that good reconstruction is achieved if the first POD coefficients are well reconstructed and even if somewhat high errors are found on the higher-order coefficients. The precursors groups and the turbulence quantities $\kappa$ and $\nu_t$ show the highest errors; this is consistent with the physical role of these fields in a circulating-fuel reactor, where precursor transport and turbulent mixing are the most spatially- and temporally-complex quantities in the coupled system, and are the fields with no direct counterpart in the input trajectories beyond their lumped, in-core-averaged value.

\begin{figure}[tp]
    \centering
    \includegraphics[width=1\linewidth]{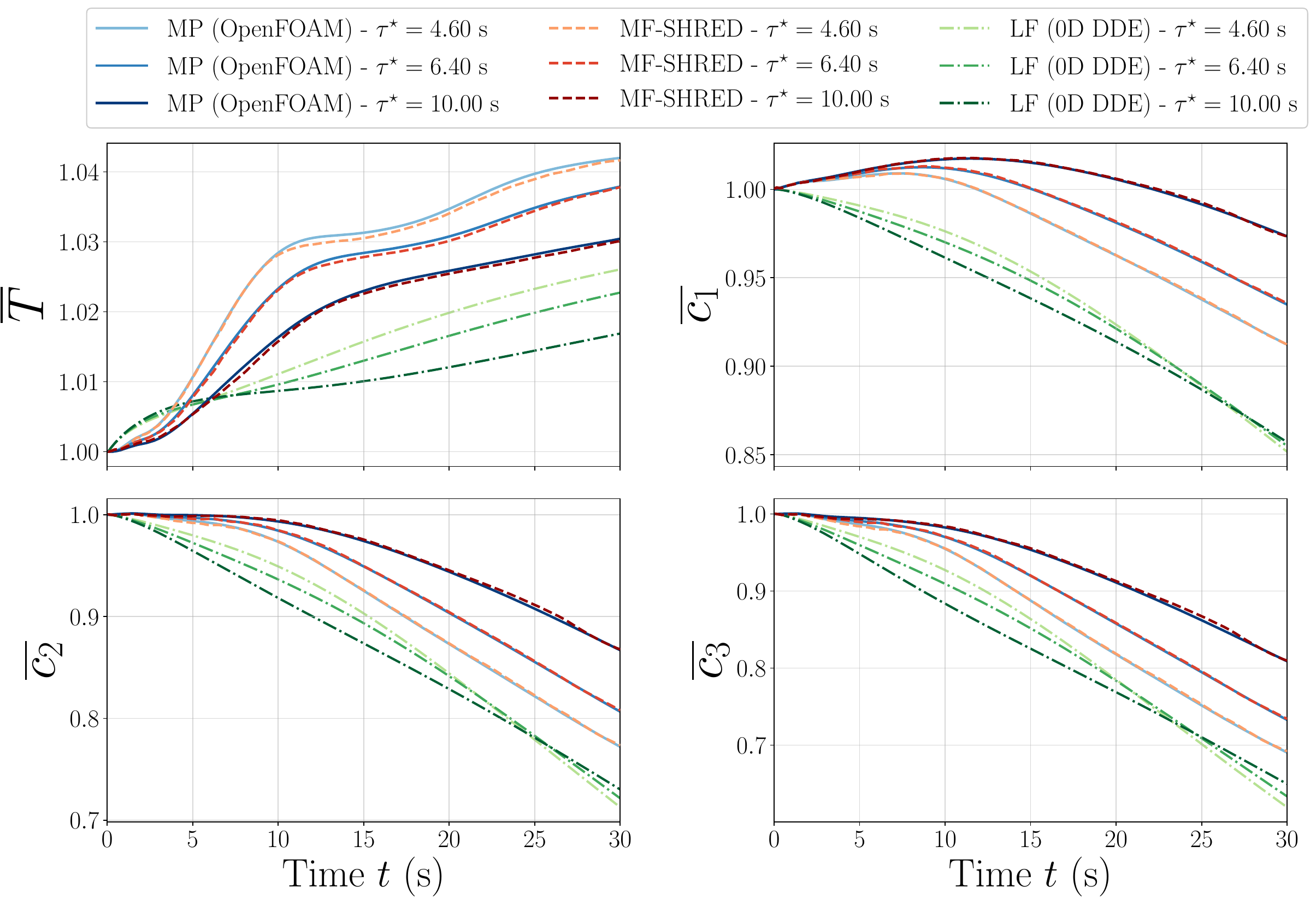}
    \caption{In-core spatial average of the fuel temperature and the first three precursor groups over time, for the three test values of $\tau^*$, comparing the high-fidelity multiphysics model, the MF-SHRED reconstruction, and the low-fidelity 0D DDE trajectory, all normalised by their initial value.}
    \label{fig: MSFR-fom_shred_lf_QoIcomparison}
\end{figure}

Figure \ref{fig: MSFR-fom_shred_lf_QoIcomparison} now compares the in-core spatial average of four representative fields (the fuel temperature and the first three precursor groups) across the three test values of $\tau^*$, contrasting the high-fidelity solution, the MF-SHRED reconstruction, and the lumped trajectory, all normalised by their initial value. MF-SHRED tracks the high-fidelity spatially-averaged response closely for all three transients, correcting the systematic deviation of the low-fidelity model, which underestimates the temperature rise and overestimates the precursor depletion rate as the transient progresses. This confirms that MF-SHRED is able to recover the high-fidelity dynamics even when the low-fidelity driving input itself is a comparatively poor approximation of the true multiphysics response.

\begin{figure}[tp]
    \centering
    \includegraphics[width=1\linewidth]{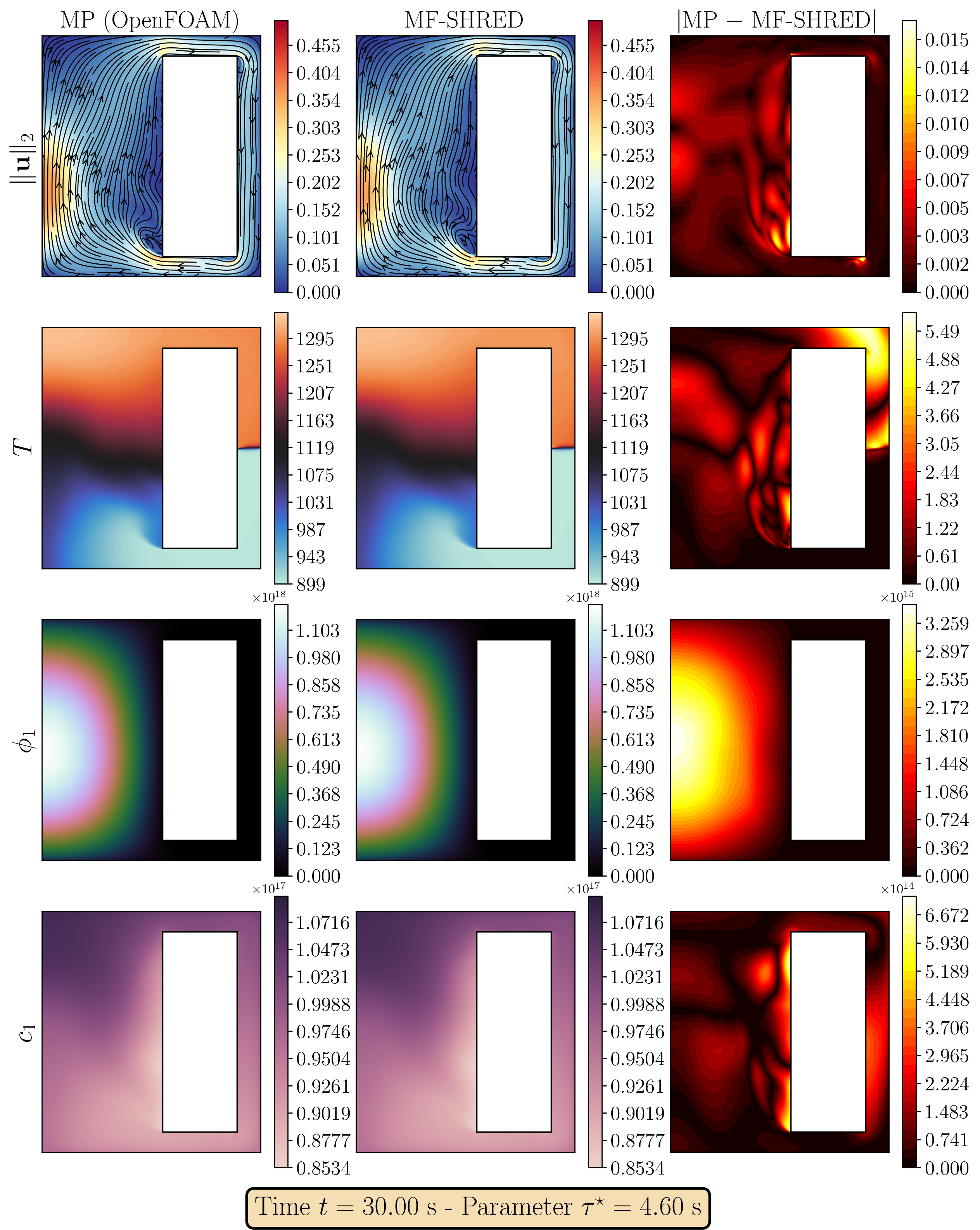}
    \caption{Spatial comparison at the final simulation time ($t = 30$ seconds) for $\tau^\star = 4.60$ s, showing the velocity magnitude, temperature, fast flux, and first precursor group: OpenFOAM solution (left), MF-SHRED reconstruction (centre), and absolute residual (right).}
    \label{fig: MSFR-spatial_snapshot_comparison_testmu_0}
\end{figure}

\begin{figure}[tp]
    \centering
    \includegraphics[width=1\linewidth]{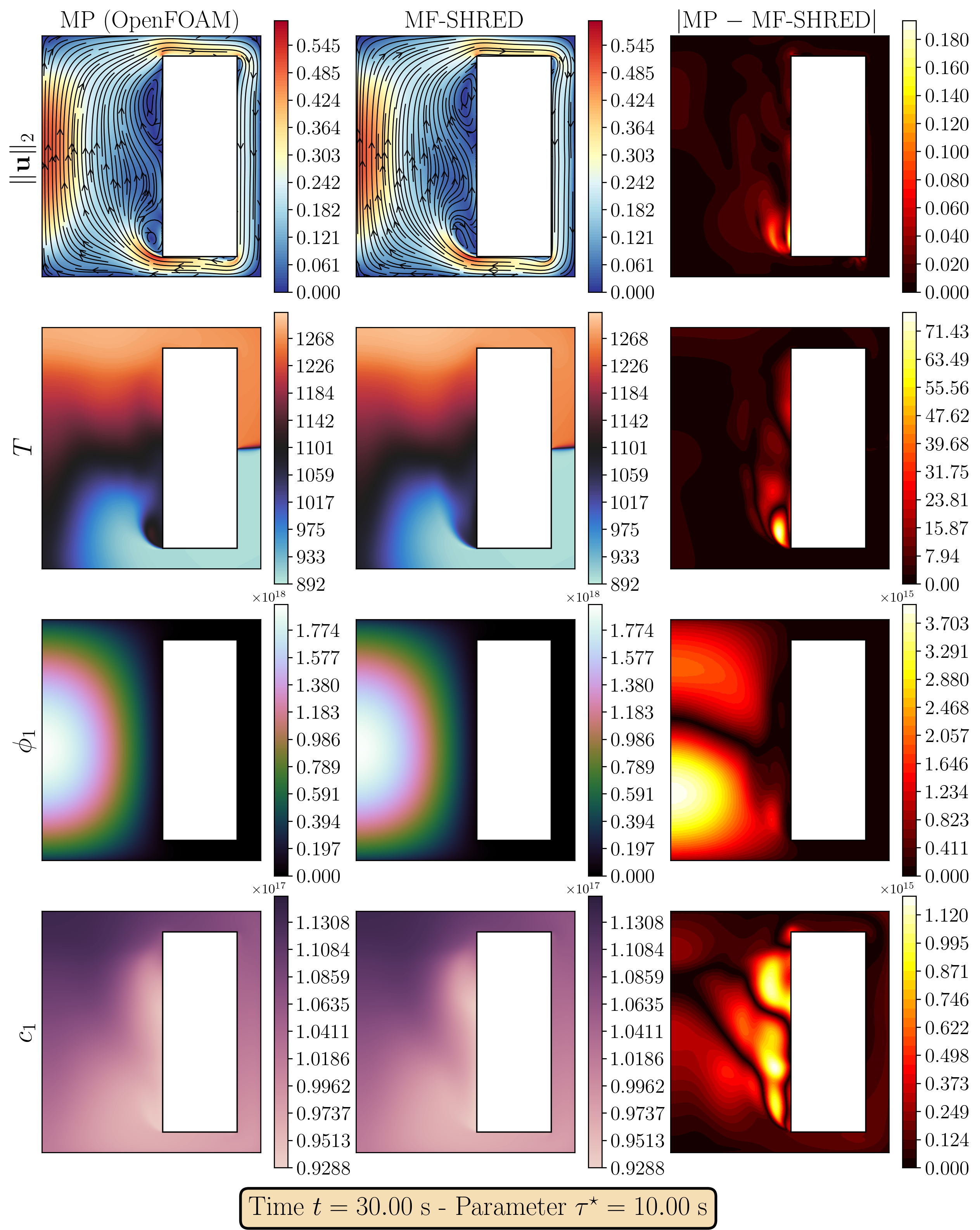}
    \caption{Spatial comparison at the final simulation time ($t = 30$ seconds) for $\tau^\star = 10$ s, showing the velocity magnitude, temperature, fast flux, and first precursor group: OpenFOAM solution (left), MF-SHRED reconstruction (centre), and absolute residual (right).}
    \label{fig: MSFR-spatial_snapshot_comparison_testmu_2}
\end{figure}

Finally, Figures \ref{fig: MSFR-spatial_snapshot_comparison_testmu_0} and \ref{fig: MSFR-spatial_snapshot_comparison_testmu_2} compare the full spatial reconstruction of the velocity magnitude, temperature, fast flux and first precursor group against the OpenFOAM solution at the final simulation time, for two test values of $\tau^\star$ (4.60 and 10 seconds). The MF-SHRED reconstruction is very close to the high-fidelity solution for all four fields; the residual fields show that the largest local errors are concentrated near the pump and heat-exchanger region and along internal flow-structure boundaries, where the spatial gradients are the largest. The residual magnitude is noticeably larger for the longer coast-down transient ($\tau^\star = 10 $ s) than for the faster one, consistent with the longer transient allowing more time for the flow and thermal fields to develop the fine-scale structures that are hardest to capture with a rank-10 POD truncation.

These results confirm that MF-SHRED achieves good performances also for the most demanding fidelity gap considered in this work: a lumped, delay-differential model that tracks only ten scalar quantities is sufficient to drive an accurate reconstruction of a 22-field, three-dimensional coupled neutronics-thermal-hydraulics multi-physics solution, achieving a mean relative error of 2.12\% on the latent dynamics and reconstructed fields that are very close from the OpenFOAM solution. Interestingly, MF-SHRED does not only reproduce the low-fidelity trajectory at high resolution: it also corrects the bias of the 0D model, tracking the true multiphysics response of the temperature and precursor fields far more closely than the low-fidelity input alone. 

\begin{figure}{tp}
    \centering
    \includegraphics[width=1\linewidth]{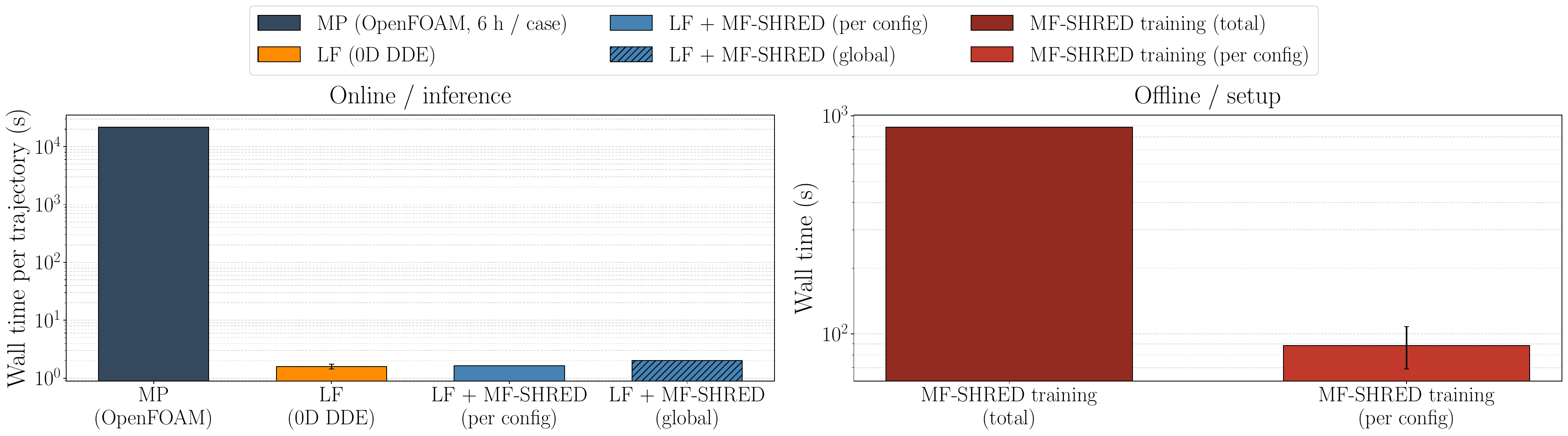}
    \caption{Computational cost comparison for the MSFR test case. Left: online wall time per trajectory for the high-fidelity solver, the low-fidelity DDE solver, and MF-SHRED, using either a single configuration or the full ensemble. Right: offline setup cost, comparing the total POD fit time against the total and per-configuration MF-SHRED training time.}
    \label{fig: MSFR-computational_cost_comparison}
\end{figure}

This is achieved at a computational cost dramatically lower than the high-fidelity solver: integrating the lumped delay-differential model with the delay-differential equation solver ($\texttt{jitcdde}$ takes on the order of 30 seconds per transient, compared to around 6 hours per transient for the coupled OpenFOAM simulation on 6 CPUs of a  workstation, resulting in a gap of roughly three-to-four orders of magnitude even before accounting for MF-SHRED (see Figure). The additional cost of MF-SHRED on top of the low-fidelity solver is comparatively negligible. Clearly, the computational advantage of MF-SHRED is most pronounced in this test case, where the low-fidelity model is a very cheap surrogate of the high-fidelity solution, and the fidelity gap is the largest of the three cases considered in this work.

 \section{Conclusions}\label{sec: concl}

This paper has investigated the use of Shallow Recurrent Decoders for multi-fidelity reduced-order modelling, extending previous works that focused exclusively on the monitoring capabilities of the architecture, i.e., the reconstruction of a high-dimensional field from sparse sensor measurements. Building on the observation that a low-fidelity model can be interpreted as a global, spatially-averaged measurement of the corresponding high-fidelity field, this work has shown that SHRED can instead be driven by the solution of a low-fidelity model, effectively replacing physical sensors with the output of a cheap surrogate that can be evaluated before a facility is built or instrumented.

The methodology has been assessed on three test cases characterised by a progressively increasing fidelity gap. On the LRA benchmark, mapping a two-group point-kinetics model to a two-group diffusion solution, MF-SHRED reconstructs the neutron fluxes and precursor concentrations with relative errors below 2\%, at a computational cost more than two orders of magnitude lower than the finite-element diffusion solver (3.57 s versus under 0.01 s per transient). On the reaction-advection-diffusion case, in which the low-fidelity lumped model discards all spatial information, MF-SHRED remains close to the POD truncation floor for all six chemical species and performs comparably to, and in some species better than, a baseline SHRED model driven by sparse high-fidelity sensors, despite having access only to spatially-averaged information; the accompanying ensemble strategy, averaging predictions over multiple random low-fidelity input subsets, further reduces the reconstruction variance at a moderate additional offline training cost. On the most demanding test case, the Molten Salt Fast Reactor, a ten-variable lumped delay-differential model is sufficient to drive an accurate reconstruction of a 22-field, three-dimensional coupled neutronics-thermal-hydraulics-CFD solution, with a mean relative error of 2.12\% on the latent dynamics; MF-SHRED is shown to also correct its systematic bias, tracking the true multi-physics response of the temperature and precursor fields more closely than the low-fidelity input alone. The computational advantage in this last case is the most striking of the three: integrating the lumped delay-differential model takes on the order of 30 seconds per transient, against several hours per transient for the coupled OpenFOAM simulation on a standard workstation, a gap of roughly three orders of magnitude even before accounting for the negligible additional cost of MF-SHRED inference.

Taken together, these results have several practical implications for the deployment of MF-SHRED in reactor engineering workflows. Because MF-SHRED can be driven by an existing low-fidelity model rather than physical sensors, it is directly applicable during the design phase of a reactor, before any instrumentation exists. This makes MF-SHRED a promising tool for early-stage digital twin development, where a plant simulator or system code can be reused as-is to drive high-fidelity state estimation without additional instrumentation cost. The two-to-three orders of magnitude speed-up demonstrated across all three cases makes previously unfeasible multi-query analyses, such as uncertainty quantification, sensitivity analysis, design optimisation, and safety-margin screening across accident scenario, feasible within the computational budget of the low-fidelity model alone, while retaining high-fidelity spatial resolution in the reconstructed output. The comparable accuracy of MF-SHRED relative to the sparse-sensor baseline indicates that, at least for the fidelity gaps considered here, a low-fidelity validated model already in use for plant control is an adequate substitute for physical instrumentation, without requiring additional sensor placement optimisation studies. In both the reaction-diffusion and MSFR cases, training an ensemble of independent MF-SHRED configurations (5 and 10 respectively) and averaging their predictions meaningfully reduced the reconstruction variance at a modest additional offline cost. For safety-relevant deployments, this ensembling is recommended as standard practice, since it also naturally provides an uncertainty estimate on the reconstructed field, which single-configuration SHRED does not. Finally, the MSFR results indicate that fields with no direct counterpart in the low-fidelity model are reconstructed with higher relative error than fields more directly related to the lumped state. Where high accuracy on these specific quantities is safety-relevant, this suggests a hybrid deployment strategy, complementing the low-fidelity-driven MF-SHRED input with a small number of local sensors on the most critical fields, rather than relying on the low-fidelity input alone.

In terms of future works, the framework should be extended to more complex, licensable reactor geometries, such as the TRIGA Mark II. Second, the multi-fidelity mapping demonstrated here between two fidelity levels can in principle be chained into a multi-step procedure, for instance, from point kinetics to diffusion to transport, or from a coarse to a fine mesh discretisation, progressively recovering higher-fidelity detail at each stage. Third, this architecture can be more easily adapted in Data Assimilation strategies like Kalman Filtering with respect to standard SHRED, combining low-fidelity global input with a small number of local sensors on fields poorly represented by the lumped model, should be investigated, particularly for turbulence and precursor-drift quantities in circulating-fuel reactors. Finally, given the promising accuracy-to-cost trade-off demonstrated across all three cases, future work should assess the integration of MF-SHRED within real-time digital twin frameworks for reactor control and monitoring, including a formal assessment of extrapolation behaviour outside the training parameter range, which is essential before any deployment in a safety-relevant context.

\color{black}
\section*{Acknowledgments}

S. Riva wishes to thank Lorenzo Loi for the help in implementing the OpenMC model for the LRA reactor.

\section*{Code and supplementary materials}  
The code and compressed data are available at: \href{github.com/ERMETE-Lab/NuSHRED}{github.com/ERMETE-Lab/NuSHRED}.

\pagebreak
\appendix
\section{Detailed Description of the Physics Models of the Test Cases}\label{app: params-testcases}

\subsection{LRA-BWR2D Reactor}\label{app: lrabench}

Two models have been implemented for this problem: a low-fidelity one based on the PK equations with two groups of precursors and a high-fidelity two-energy group diffusion model. The PK model, which has been solved in Python, using the \texttt{solve\_ivp} solver from the \texttt{scipy} library, reads as follows:
\begin{subequations}
    \label{eqn: pk}
    \begin{empheq}[left=\empheqlbrace]{align}
        \der{P}{t} &= \frac{\rho(t; \boldsymbol{\mu})-\beta}{\Lambda}P + \sum_{j=1}^J\lambda_jc_j \label{eqn: pk_power} \\
        \der{c_j}{t} &= \frac{\beta_j}{\Lambda}P - \lambda_jc_j \label{eqn: pk_precursor}
    \end{empheq}
\end{subequations}
where $P$ is the reactor power, $c_j$ is the concentration of the $j-$th precursors group, $\rho(t; \boldsymbol{\mu})$ is the reactivity, $\beta$ is the total delayed neutron fraction, $\beta_j$ is the delayed neutron fraction of the $j-$th group, $\Lambda$ is the neutron generation time, and $\lambda_j$ is the decay constant of the $j-$th precursors group; in the end, $\boldsymbol{\mu}$ represents a vector of parameters specific to the undergoing transient starting from criticality conditions. The multi-group diffusion equation, with $\phi_g$ as the neutron flux of the $g-$th group and $c_j$ the $j-$th precursors groups, reads:

\begin{subequations}
    \begin{empheq}[left=\empheqlbrace]{align}
        \frac{1}{v_g}\dpart{\phi_g}{t} &= \nabla\cdot(D_g\nabla \phi_g)
        -\left(\Sigma_{a,g}(t;\boldsymbol{\mu})
        +\sum_{g'\neq g}\Sigma_{s,g\rightarrow g'}\right)\phi_g
        \quad +\sum_{g'\neq g}\Sigma_{s,g'\rightarrow g}\phi_{g'}+\chi_g S_g \\
        \dpart{c_j}{t} &= \frac{\beta_j}{k_{\text{eff}}}\sum_{g}\nu_g\Sigma_{f,g}\phi_{g} - \lambda_jc_j
    \end{empheq}
    \label{eqn: mg-neutron-diffusion}
\end{subequations}
with $S_g = \frac{1-\beta}{k_{\text{eff}}}\sum_{g'}\nu_{g'}\Sigma_{f,g'}\phi_{g'}+\sum_{j}\lambda_jc_j$ being the neutron source term and $\Sigma_{x,g}$ as the group constant for generic reaction $x$ of energy group $g$. The system starts from initial critical conditions and the boundary conditions are zero flux on the external boundary of the domain (top and right sides) and reflective on the internal boundaries (left and bottom sides), see Figure \ref{fig: lra-domain-region}. This system of partial differential equations has been discretised using the finite element method, provided by the open-source package \textit{dolfinx} (version 0.10.0) \cite{BarattaEtal2023, ScroggsEtal2022, BasixJoss, AlnaesEtal2014} and updating the code developed by the authors in \cite{LOI2024113480} (the original code can be found at \href{https://github.com/ERMETE-Lab/MP-OFELIA}{github.com/ERMETE-Lab/MP-OFELIA} whereas the update is part of the NuSHRED repository, companion of this work). Transient behaviour is triggered by modifying the absorption cross section $\Sigma_{a,2}(\vec{x}, t; \boldsymbol{\mu})$ of the thermal group in the rod region (ID = 60), $\mathbf{x}\in\Omega_{rod}$, from the nominal value $\Sigma_{a,2}^{(0)}(\vec{x})$ as follows: $\Sigma_{a,2}(\vec{x}, t; \boldsymbol{\mu}) = \Sigma_{a,2}^{(0)}(\vec{x})\cdot f(t; \boldsymbol{\mu})$. The function $f$ depends on time and on parameters identifying the specific transient under which the system undergoes (such as the height of a step or the end time of a ramp). A similar input condition is given to the PK model through reactivity $\rho(t, \boldsymbol{\mu})$. 

Three different profiles for the absorption cross section have been studied, namely ramp, sinusoidal, and trapezoidal signals. The general form of these functions is given below, dependent on different parameters $\mu_1\in[0.97, 1.03]$, $\mu_2\in[0.2, 0.8]$, $\mu_3\in[0.2, 0.75]$ and $\mu_4\in[0.2, 0.79]$: the first one represents the final value of the modified cross section $\pm 3\%$, the second and fourth ones are the end time of the rising ramp, whereas the third one is the period of the oscillation.
\begin{subequations}
    \begin{align}
         f_{\text{ramp}}(t; \boldsymbol{\mu}) &= 
         \left\{
            \begin{array}{cc}
                1+(\mu_1 - 1)\cdot \frac{t}{\mu_2} & 0\leq t\leq \mu_2 \\ 
                \mu_1 & t>\mu_2
            \end{array}
         \right.
         \\
         f_{\text{sin}}(t; \boldsymbol{\mu}) &= 1 + \frac{\mu_1 - 1}{2}\cdot \left(1 - \cos\left(\frac{2\pi t}{\mu_3}\right)\right)\\
         f_{\text{trapz}}(t; \boldsymbol{\mu}) &= 
         \left\{
            \begin{array}{cc}
                1 + (\mu_1 - 1) \cdot \frac{t}{\mu_4} & 0\leq t\leq \mu_4 \\ 
                \mu_1 & \mu_2<t\leq 0.8 \cdot T_{final}\\
                1 + (\mu_1 - 1) \cdot \frac{T_{final} - t}{T_{final} \cdot (1-0.8)} & t> 0.8 \cdot T_{final}
            \end{array}
         \right.
        \end{align}
    \end{subequations}
    
The equivalent functions for the reactivity profiles are not reported for the sake of brevity (see the NuSHRED Github repository, if interested).

\subsection{Reaction-Diffusion-Advection of species}\label{app: reactdiff}

Complementary to the previous linear multi-group diffusion benchmark, in which an almost exact linear map exists between the two fidelity levels, a second test case is introduced to assess multi-fidelity state reconstruction in a more challenging setting. The governing equation retains the diffusion--reaction structure of the neutron flux model, but it is here extended to a coupled system of non-linear advection--diffusion--reaction PDEs for $I$ chemical species. Systems of this kind are commonly employed to describe the transport of contaminants, reactants, or radioactive isotopes in fluid flows. In this context, the high-fidelity (HF) model resolves the spatial distribution of each species on $\Omega\subset\mathbb{R}^2$, whereas the low-fidelity (LF) model provides only domain-averaged trajectories obtained from a lumped ODE formulation. The main modelling discrepancy between the two fidelities therefore arises from spatial heterogeneity and from the closure of non-linear reaction terms.

Let $c_i(\vec{x},t)$ denote the concentration of species $i$ and $\vec{c}=[c_0,\,c_1,\,\ldots,\,c_{I-1}]^{\mathsf T}\in\mathbb{R}^I$ the associated state vector. From a local mass balance, each component satisfies
\begin{equation}\label{eq:rdadv-pde}
\dpart{c_i}{t} + \nabla \cdot (\vec{u}\, c_i) - D_i \Delta c_i = r_i(\vec{c}) + s_i(\vec{x},t),
\qquad \vec{x}\in\Omega,\; t\in\mathcal{T},
\end{equation}
where $\vec{u}(\vec{x})$ is a prescribed solenoidal velocity field, the species are treated as passive scalars in the flow, $D_i>0$ is the diffusion coefficient of species $i$, $s_i$ is an external source, and $r_i(\vec{c})$ collects the internal reactions. When $r_i$ is linear in $\vec{c}$, the system reduces to a coupled generalisation of the Bateman equations. To introduce a controlled non-linearity, mass-action terms of second order are also retained:
\begin{equation}\label{eq:rdadv-react}
    r_i(\vec{c}) = \vec{R}_i^{\mathsf T}\vec{c} - \sum_{j,k=0}^{I-1} Q_{ijk}\, c_j c_k,
\end{equation}
where $Q_{ijk}$ denotes the quadratic reaction tensor adopted consistently in both fidelities. When $Q_{ijk}=Q_{ikj}$, \eqref{eq:rdadv-react} can equivalently be written as $\vec{c}^{\mathsf T}\vec{R}_{2,i}\vec{c}$ with $[\vec{R}_{2,i}]_{jk}=-Q_{ijk}$. More general reaction laws could be considered in principle; however, the present benchmark focuses on the closure error induced by \eqref{eq:rdadv-react}. The HF initial-boundary value problem is closed by non-uniform initial profiles and homogeneous Neumann (zero-flux) boundary conditions on $\partial\Omega$. From the numerical point of view, \eqref{eq:rdadv-pde}--\eqref{eq:rdadv-react} is discretised in space with the finite element method using the open-source library dolfinx (version~0.10.0), and integrated in time with an implicit Euler scheme and a semi-implicit treatment of the quadratic terms\footnote{The non-linear contribution $-\sum_{j,k} Q_{ijk}\, c_j c_k$ is evaluated at time $t_k$ as $-\sum_{j,k} Q_{ijk}\, c_j(t_{k-1})\, c_k(t_k)$.}.

The HF model follows from a local conservation balance over an infinitesimal control volume. Conversely, the LF model is obtained by enforcing a global balance over $\Omega$:
\begin{equation}\label{eq:rdadv-lumped-balance}
    \der{C_i}{t} + J_+ - J_- = R_i(\vec{C}) + S_i(t),
\end{equation}
where $C_i(t)$ is the domain average of species $i$, $R_i$ and $S_i$ are the lumped reaction and source terms, and $J_\pm$ denote the inward and outward fluxes across $\partial\Omega$. In the present setting, homogeneous Neumann conditions and a solenoidal velocity field with $\vec{u}\cdot\vec{n}=0$ on $\partial\Omega$ imply $J_+=J_-=0$, so that \eqref{eq:rdadv-lumped-balance} reduces to a closed ODE system in $\vec{C}(t)$ once $R_i$ and $S_i$ are specified.

It is worth mentioning that, unlike the linear diffusion benchmark, the map between HF and LF descriptions is not exact in general: spatially varying sources, advection, and quadratic reactions introduce a model discrepancy that must be explicitly accounted for when LF trajectories are used as inputs to multi-fidelity SHRED. The remainder of this section derives the LF model from \eqref{eq:rdadv-pde} and states the assumptions underlying the closure.

\subsubsection{From the PDE model to the lumped ODE formulation}\label{sec:rdadv-closure}

The domain average of species $i$ is defined as
\begin{equation}\label{eq:rdadv-average}
C_i(t) = \frac{1}{|\Omega|} \int_\Omega c_i(\vec{x}, t) \, d\vec{x}.
\end{equation}
Integrating \eqref{eq:rdadv-pde} over $\Omega$ and invoking the divergence theorem yields
\begin{equation}\label{eq:rdadv-integrated}
    \int_\Omega \left(\dpart{c_i}{t} + \nabla \cdot (\vec{u}\, c_i) - D_i \Delta c_i - r_i(\vec{c}) - s_i\right) \, d\vec{x} = 0.
\end{equation}
Each term in \eqref{eq:rdadv-integrated} is analysed below to connect \eqref{eq:rdadv-pde} with the lumped dynamics $\der{C_i}{t}=R_i(\vec{C})+S_i(t)$.

\begin{itemize}
    \item \textit{Time derivative.} Since $\Omega$ is fixed in time, differentiation and integration can be interchanged, so that
    \begin{equation}
        \frac{1}{|\Omega|}\int_\Omega \dpart{c_i}{t} \, d\vec{x} = \der{C_i}{t}.
    \end{equation}
    \item \textit{Advection.} The divergence theorem gives
    \begin{equation}
        \int_\Omega \nabla \cdot (\vec{u}\, c_i) \, d\vec{x} = \int_{\partial \Omega} (\vec{u}\cdot\vec{n})\, c_i \, dS,
    \end{equation}
    where $\vec{n}$ is the outward unit normal on $\partial\Omega$. Assuming impermeability of the boundary to the flow, i.e.\ $\vec{u}\cdot\vec{n}=0$ on $\partial\Omega$, the advection term does not contribute to the domain-averaged balance.
    \item \textit{Diffusion.} Integration by parts yields
    \begin{equation}
        \int_\Omega D_i \Delta c_i \, d\vec{x} = D_i \int_{\partial \Omega} \nabla c_i \cdot \vec{n} \, dS,
    \end{equation}
    which vanishes under the homogeneous Neumann condition $\nabla c_i\cdot\vec{n}=0$ on $\partial\Omega$.
    \item \textit{Source term.} Defining the lumped source as the domain average
    \begin{equation}
        S_i(t) = \frac{1}{|\Omega|} \int_\Omega s_i(\vec{x}, t) \, d\vec{x},
    \end{equation}
    the contribution of $s_i$ to $\der{C_i}{t}$ is exact. When $s_i(\vec{x},t)=\lambda_i(t)\,\widetilde{S}_i(\vec{x})$, this reduces to $S_i(t)=\lambda_i(t)\,\overline{\widetilde{S}_i}$, where the overbar denotes spatial averaging.
    \item \textit{Linear reactions.} If $r_i(\vec{c})=\vec{R}_i^{\mathsf T}\vec{c}$, the averaged reaction rate is also exact:
    \begin{equation}
        R_i(\vec{C}) = \frac{1}{|\Omega|} \int_\Omega \vec{R}_i^{\mathsf T}\vec{c} \, d\vec{x} = \vec{R}_i^{\mathsf T}\vec{C}(t).
    \end{equation}
    \item \textit{Quadratic reactions.} For $r_i^{(2)}(\vec{c})=-\sum_{j,k} Q_{ijk}\, c_j c_k$, a Reynolds decomposition is applied to each species,
    \begin{equation}
        c_j(\vec{x}, t) = C_j(t) + c_j'(\vec{x}, t), \qquad \int_\Omega c_j'(\vec{x}, t)\, d\vec{x} = 0.
    \end{equation}
    Expanding $c_j c_k$ and averaging over $\Omega$ gives
    \begin{equation}
        \frac{1}{|\Omega|}\int_\Omega c_j c_k \, d\vec{x}
        = C_j C_k + \frac{1}{|\Omega|}\int_\Omega c_j' c_k' \, d\vec{x},
    \end{equation}
    because the cross terms integrate to zero. Therefore,
    \begin{equation}
        R_i^{(2)}(\vec{C}) = \frac{1}{|\Omega|}\int_\Omega r_i^{(2)}(\vec{c})\, d\vec{x}
        = -\sum_{j,k} Q_{ijk}\, C_j C_k + \frac{1}{|\Omega|}\sum_{j,k} Q_{ijk}\int_\Omega c_j' c_k' \, d\vec{x}.
    \end{equation}
    The LF model neglects the fluctuation integral in the last expression. This closure assumption is generally valid only when spatial gradients are sufficiently small; otherwise, it represents the main source of discrepancy between the HF and LF models.
\end{itemize}

Collecting the previous results, and assuming that i) reaction terms are closed at the domain average, ii) no net flux crosses $\partial\Omega$, and iii) quadratic fluctuation integrals are negligible, the lumped ODE system reads
\begin{equation}\label{eq:rdadv-ode}
    \der{C_i}{t} = S_i(t) + R_i(\vec{C})
    = S_i(t) + \vec{R}_i^{\mathsf T}\vec{C}(t) - \sum_{j,k=0}^{I-1} Q_{ijk}\, C_j(t)\, C_k(t).
\end{equation}
In particular, the LF trajectories used in the multi-fidelity SHRED experiments are obtained by time integration of \eqref{eq:rdadv-ode}. Even though the derivation above clarifies the assumptions under which HF and LF dynamics coincide, future investigation may focus on improved closures for advection-dominated regimes and on spatially localised sources, for which the gap between \eqref{eq:rdadv-pde} and \eqref{eq:rdadv-ode} is expected to be more pronounced.

\subsubsection{Parameters for the test case}\label{sec:rdadv-params}

The benchmark is defined on the unit square $\Omega=(0,1)^2$ with $I=6$ chemical species. Both the HF and LF models are evaluated on the same temporal grid.

The HF problem is discretised on a uniform triangular mesh with $50$ cells per direction ($N=50$), using piecewise-linear Lagrange finite elements for each species. Homogeneous Neumann conditions are imposed on $\partial\Omega$. Time integration employs an implicit Euler scheme with time step $\Delta t=5\times 10^{-3}$ up to $T=10$; snapshots are stored every $10$ steps.

Species are advected by a prescribed single-gyre velocity field,
\begin{equation}
    \vec{u}(x,y) = u_\mathrm{scale}
    \begin{bmatrix}
        -\sin(\pi x)\cos(\pi y) \\
        \cos(\pi x)\sin(\pi y)
    \end{bmatrix},
    \qquad u_\mathrm{scale}=5,
\end{equation}
which is solenoidal and tangent to $\partial\Omega$, so that $\vec{u}\cdot\vec{n}=0$ on the boundary. The diffusion coefficients are species-dependent and fixed to
$D_0=2\times 10^{-3}$, $D_1=10^{-3}$, $D_2=4\times 10^{-3}$, $D_3=2\times 10^{-3}$, $D_4=3\times 10^{-3}$, and $D_5=1.5\times 10^{-3}$.

Regarding reaction kinetics, the linear contribution in \eqref{eq:rdadv-react} is specified by the rate vectors $\vec{R}_i$ (reference case $\text{linear\_scale}=1$), which redistribute mass among the first five species, with $c_5$ acting as a terminal pool. Columns $0$--$2$ of the underlying coupling matrix are scaled uniformly by the parameter $\text{linear\_scale}$, preserving the column balance among species $0$--$4$. The quadratic network contains nine elementary mass-action steps, including self-reactions, e.g., $c_0+c_0\to c_1$, and cross-species couplings, e.g., $c_1+c_2\to c_0$; the full tensors are reported in the companion GitHub repository.

External forcing is imposed in separable form,
\begin{equation}\label{eq:rdadv-source}
    s_i(\vec{x},t)=\lambda_i(t)\,\widetilde{S}_i(\vec{x}),
\end{equation}
where $\widetilde{S}_i$ is a fixed spatial profile and $\lambda_i(t)$ is a time-dependent amplitude. Species $1$, $3$, and $5$ have $\widetilde{S}_i\equiv 0$; for the remaining source species,
\begin{equation}\label{eq:rdadv-source-spatial}
    \widetilde{S}_i(\vec{x})
    = \bigl|\sin\!\bigl(\omega_i\,(x-x_{0,i})\bigr)\,
           \sin\!\bigl(\omega_i\,(y-y_{0,i})\bigr)\bigr|,
    \qquad i\in\{0,2,4\},
\end{equation}
with $(\omega_i,x_{0,i},y_{0,i})=(2.0,\,0.15,\,0.50)$ for $i=0$, $(1.5,\,0.75,\,0.25)$ for $i=2$, and $(1.0,\,0.50,\,0.75)$ for $i=4$. The temporal pulses share the parametric centre $t_\mathrm{peak}$ and width $w$ (with $w\leftarrow\max(w,\,0.05)$ in the implementation), but differ in shape:
\begin{align}
    \lambda_0(t) &= \exp\!\left(-\tfrac{1}{2}\left(\frac{t-t_\mathrm{peak}}{w}\right)^{\!2}\right),
    \label{eq:source-lambda0}\\[4pt]
    \lambda_2(t) &=
    \begin{cases}
        \sin^2\!\left(\dfrac{\pi}{4}\,\dfrac{t-(\tfrac{3}{2}\,t_\mathrm{peak}-w)}{w}\right),
        & \tfrac{3}{2}\,t_\mathrm{peak}-w \le t \le \tfrac{3}{2}\,t_\mathrm{peak}+w,\\[6pt]
        0, & \text{otherwise},
    \end{cases}
    \label{eq:source-lambda2}\\[4pt]
    \lambda_4(t) &=
    \begin{cases}
        0, & t < \tfrac{5}{2}\,t_\mathrm{peak}-w,\\[4pt]
        \left(\dfrac{t-(\tfrac{5}{2}\,t_\mathrm{peak}-w)}{w}\right)^{\!2},
        & \tfrac{5}{2}\,t_\mathrm{peak}-w \le t \le \tfrac{5}{2}\,t_\mathrm{peak},\\[6pt]
        \exp\!\left(-\dfrac{t-\tfrac{5}{2}\,t_\mathrm{peak}}{w}\right),
        & t > \tfrac{5}{2}\,t_\mathrm{peak},
    \end{cases}
    \label{eq:source-lambda4}
\end{align}
where $\lambda_1=\lambda_3=\lambda_5\equiv 0$. The corresponding lumped source in \eqref{eq:rdadv-ode} is the integral average. Initial concentrations are non-uniform trigonometric or Gaussian profiles on $\Omega$; the LF model is initialised from their domain averages, in the same way as the source amplitudes.

Each dataset realisation is labelled by $\boldsymbol{\mu}=(\text{linear\_scale},\, t_\mathrm{peak},\, w)\in\mathcal{D}$, where $t_\mathrm{peak}$ and $w$ coincide with $\text{source\_t\_peak}$ and $\text{source\_pulse\_width}$ in the code. The parametric grid is the Cartesian product of
\begin{center}
\begin{tabular}{lll}
    $\text{linear\_scale}$ & $\in\{0.5,\,0.67,\,\ldots,\,2.0\}$ & ($10$ values, uniform in $[0.5,2]$), \\
    $t_\mathrm{peak}$ & $\in\{0.5,\,1.125,\,\ldots,\,3.0\}$ & ($5$ values, uniform in $[0.5,3]$), \\
    $w$ & $\in\{0.2,\,0.4,\,\ldots,\,2.0\}$ & ($10$ values, uniform in $[0.2,2]$),
\end{tabular}
\end{center}
yielding $500$ paired HF/LF trajectories. For each case, the LF ODE \eqref{eq:rdadv-ode} is integrated with the same pulse shapes and scaled linear kinetics as the HF run.

\subsection{Molten Salt Fast Reactor Reactors} \label{app: msfr-gov-eqn}

\subsubsection{Multi-Physics Model}

\begin{figure}[tp]
  \centering
  \includegraphics[width=0.6\linewidth]{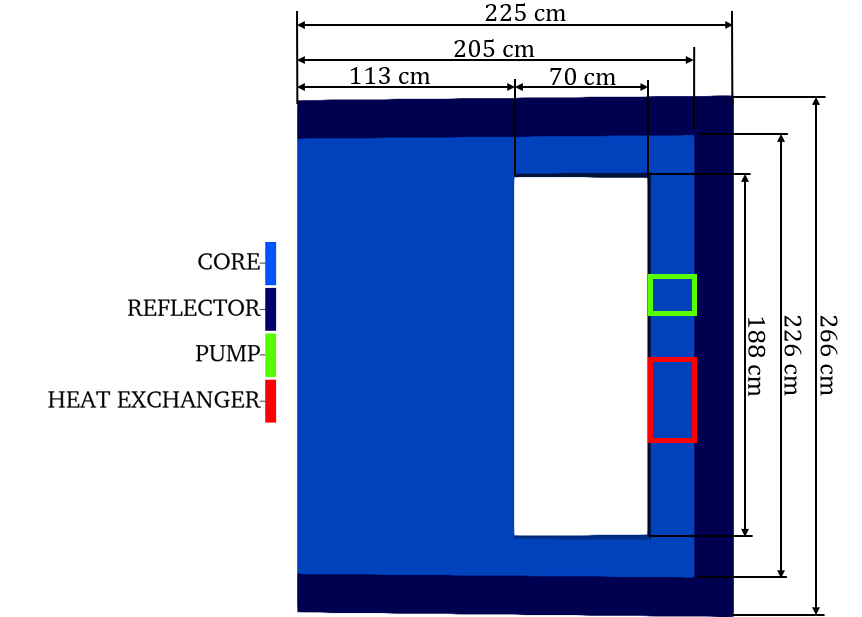}
  \caption{OpenFOAM simulation domain with the main geometric dimensions and the primary loop components. The geometry refers to a 2D axisymmetric wedge of the EVOL geometry of the European MSFR design, and includes molten salt fuel (light blue), the Hastelloy reflector (dark blue), the primary pump (green) and the heat exchanger with the intermediate cycle (red). The blank hole represents the solid salt fertile blanket, not simulated in the present model.}
  \label{fig: evol-geom}
\end{figure}

The Molten Salt Fast Reactor \cite{GenIV-RoadMap} is an innovative reactor concept in which the fuel is liquid and homogeneously mixed with the molten salt thermal carrier. Moreover, there is a strong coupling between the different physics, even more so than other nuclear reactor configurations, the mathematical modelling of such system requires suitable tools to integrate different physics together, starting from the neutronics and thermal-hydraulics \cite{aufiero2014development, CERVI2019379}.

The governing equations of the MSFR consist of the Navier-Stokes equations with the energy balance, including turbulence modelling with RANS (Realizable $\kappa-\epsilon$ model) and the Boussinesq approximation for density variations, the multi-group neutron diffusion and the precursors advection-diffusion: for completeness, this section reports the governing equations and briefly discusses their coupling. The equations that will be presented below are related to the transient problem, the initial condition is always the critical state from which the accident starts. In order to obtain the steady initial condition a $k$-eigenvalue calculation has been previously before. Interested readers can refer to \cite{aufiero2014development, CERVI2019379}. \\
The Navier-Stokes equations for the MSFR reads \cite{versteeg2007introduction}:
\begin{subequations}
\begin{align}
\nabla\cdot \vec{u} =& 0 \label{eq: mass} \\
\frac{\partial\vec{u}}{\partial t} + (\vec{u}\cdot \nabla)\vec{u} =& +\nabla\cdot [\nu_{\text{eff}}(\nabla\vec{u} + (\nabla\vec{u})^{T})] \nonumber \\
&- \frac{1}{\rho} \nabla p + [1 - \beta_{T}(T - T_{0})] \vec{g}  \label{eq: velocity} \\
\frac{\partial T}{\partial t} + \nabla \cdot \left(\vec{u} T\right)  =& \nabla\cdot (\alpha_{\text{eff}}\nabla T) + \frac{q'''}{\rho c_{p}} \label{eq: energy}
\end{align}
\end{subequations}
given $\vec{u}$ as the velocity, $\rho$ as the density, $p$ as the pressure, $\nu_{\text{eff}}$ as the effective kinematic viscosity (accounting for turbulence modelling, i.e. $\nu+\nu_T$), $T$ as the temperature, $\beta_T$ as the thermal expansion, $\vec{g}$ as the gravity acceleration, $\alpha_{\text{eff}}$ as the effective thermal diffusivity (accounting for turbulence modelling, i.e. $\frac{\nu}{Pr}+\frac{\nu_T}{Pr_T}$, with $Pr$ stands for the Prandtl number), $q'''$ as the power density, which includes both neutron fission and delayed heat sources, and $c_p$ as the specific heat capacity. The PDE system is closed by suitable boundary conditions, in particular no slip condition for the velocity and adiabatic for the temperature on the outer boundary. This set of equations shows the strong coupling between the two different physics; in particular, the source term of the energy equation directly depends on the neutron fluxes, i.e.
\begin{equation}
    q''' = (1-\beta_{h})\sum_{g}\bar{E}_{f,g}\Sigma_{f,g}\phi_{g} + \sum_{i}\lambda_{h,i}d_{i}
    \label{eqn: volumetric-heat-source}
\end{equation}
with $\phi_g$ being the neutron flux of energy group $g$, $\beta_h$ being the fraction of delayed heat source, $\bar{E}_{f,g}$ being the average energy released by fission in group $g$, $\Sigma_{f,g}$ being the fission cross section \cite{DuderstadtHamilton} of group $g$, $\lambda_{h,i}$ being the  decay constant of delayed decay group $d_i$.

Neutronics is governed by the Multi-Group Diffusion Equation \cite{DuderstadtHamilton, aufiero2014development}:
\begin{equation}	
    \frac{1}{v_{g}} \frac{\partial\phi_{g}}{\partial t} - \nabla\cdot (D_{n,g} \nabla \phi_{g}) + \Sigma_{r,g} \phi_{g} = S_{g} 
\label{eqn: diffusion-mg}
\end{equation}
with $D_{n,g}$ being the neutron diffusion coefficient of group $g$ and $\Sigma_{r,g}$ being the removal cross-section accounting for all the reactions which decrease the number of neutrons of group $g$, i.e. absorption $\Sigma_{a,g}$ and infra-group scattering $\Sigma_{s,g \to g'}$:
\begin{equation}
    \Sigma_{r,g}^0 =  \left(\Sigma_{a,g}^0+ \sum_{ g' \neq g }\Sigma_{s,g \to g'}^0 \right)
\end{equation}
where the superscript $0$ refers to the values at the reference temperature. In fact, the cross section are strongly affected by the neutron energy due to the Doppler broadening effect and the variation in density, i.e. 
\begin{equation}
    \Sigma_{r,g} = \left( \Sigma_{r,g}^{0} + A_{r,g}^{0}\ln\frac{T}{T_{0}^{\Sigma}}\right)\cdot \frac{1-\beta_{T}(T - T_{0})}{1 - \beta_{T}(T_{0}^{\Sigma} - T_{0})}
    \label{eqn: xs-mg}
\end{equation}
with $A_{r,g}^{0}$ as a fitting coefficient for the log-law and $T_{0}^{\Sigma}$ is the reference temperature at which the reference group cross section $\Sigma_{r,g}^{0}$ is calculated. In the end, the neutron diffusion equations have a source term $S_g$ accounting for the production of neutrons in the group energy $g$ due to prompt fission, decay of precursors and infra-group scattering:
\begin{subequations}
\begin{align}
    S_{g} &= (1-\beta) \chi_{p,g}\,S_{p,g} +\chi_{d,g}\,S_{d} + S_{s,g}\\
      S_{p,g} &= \sum_{g'}\bar{\nu}_{g'}\Sigma_{f,g'}\phi_{g'} \\
      S_{d}   &=\sum_{k}\lambda_{k}c_{k} \\
      S_{s,g} &= \sum_{g' \neq g}\Sigma_{s,g' \to g} \phi_{g'}
\end{align}
\end{subequations}
with $\chi_{p,g}$ and $\chi_{d,g}$ as the fission spectrum of prompt and delayed neutrons and $\bar{\nu}_{g'}$ as the average number of neutrons produced in group $g'$. The diffusion equations for the neutron flux is closed by albedo boundary conditions.

When dealing with transient scenarios, the production of neutrons is split into prompt (due to fission) and delayed (due to decay): in the nuclear reactors world, this latter contribution is described by the group precursors $c_k$ which are governed by an advection-diffusion equation in the salt (closed by homogenous Neumann condition at the outer boundary:
\begin{equation}
\begin{split}
\frac{\partial c_{k}}{\partial t} + \nabla\cdot (\vec{u} c_{k}) =& +\nabla\cdot (D_{\text{eff}} \nabla c _{k}) - \lambda_{k} c _{k} + \beta_{d,k}\sum_{g} \bar{\nu}_{g} \Sigma_{f,g}\phi_{g}
\end{split}
\label{eqn: precursors}
\end{equation}
with $D_{\text{eff}}$ the effective diffusion coefficient (accounting for turbulence modelling, i.e. $\frac{\nu}{Sc}+\frac{\nu_T}{Sc_T}$, with $Sc$ stands for the Schmidt number) and $\lambda_k$ the decay constant of delayed neutron precursor group $k$. In the end, the precursors contribute also to the energy equation \eqref{eq: energy}, where they are grouped together into the decay heat group $d_i$
\begin{equation}
\begin{split}
    \frac{\partial d_{i}}{\partial t} + \nabla\cdot(\vec{u}d_{i}) =& + \nabla\cdot(D_{\text{eff}}\nabla d_{i}) - \lambda_{h,i}d_{i} + \beta_{h,l}\sum_{g}\bar{E}_{f,g}\Sigma_{f,g}\phi_{g}
\end{split}
\label{eqn: decay-heat-prec}
\end{equation}

\subsubsection{Lumped 0D Model}\label{app: msfr-0d}

The low-fidelity model of the MSFR couples one-group point kinetics with eight circulating delayed-neutron precursor groups, a lumped fuel energy balance, and a linear fuel-temperature reactivity feedback:
\begin{subequations}
\begin{align}
    \frac{dP}{dt} &= \frac{\rho(t)-\beta}{\Lambda}P(t) + \sum_{i=1}^8\lambda_iC_i(t) \\
    \frac{dC_i}{dt} &= \frac{\beta_i}{\Lambda}P(t) - \lambda_i C_i(t) - \frac{C_i(t)}{\tau_f} + \frac{C_i(t-\tau_{ec})}{\tau_f}e^{-\lambda_i \tau_{ec}}, \quad i=1,\dots,8 \\
    \frac{dT_F}{dt} &= \frac{P(t)}{M_F c_F} - \frac{2\Gamma}{M_F}  (T_F(t) - T_{F,in}) \\
    \rho(t) &= \rho_{ext} + \alpha_F(T_F(t) - T_F^0)
\end{align}
\end{subequations}
where $P(t)$ is the core thermal power, $C_i$ is the density of the $i$-th precursor group, $T_F$ is the average core fuel temperature, $\beta_i$ and $\lambda_i$ are the delayed-neutron fraction and decay constant of group $i$, $\Lambda$ is the prompt neutron generation time, $\alpha_F$ is the fuel temperature reactivity coefficient, $M_F$ is the core fuel inventory, $c_F$ is the fuel specific heat, and $\Gamma$ is the fuel mass flow rate. 

Compared to solid-fuel lumped models, the MSFR must also consider the fact that the fuel is liquid and circulates outside the core through the primary loop. Therefore, a fraction of the delayed-neutron precursors decays outside the active region and re-enters the core only after the external-loop transit time $\tau_{ec}$, with the corresponding decay factor $\exp(-\lambda_i\tau_{ec})$. This transport delay makes the MSFR lumped model a system of delay differential equations (DDE) rather than a standard ODE system: to solve it, the \texttt{jitcdde} integrator is used \cite{SHAMPINE2001441}.

The transient considered in this work is an Unprotected Loss Of Flow (ULOFF), in which the primary pump coasts down exponentially without scram $u(t)=u_0\exp(-t/\tau)$ being $\tau$ the coast-down time constant. As the pump slows, the fuel residence time $\tau_f$ and the precursor transit time $\tau_{ec}$ scale inversely with $u(t)$, the mass flow rate $\Gamma$ scales proportionally with $u(t)$ and the heat exchanger effectiveness is modelled as increasing with decreasing flow, so that the core inlet temperature $T_{F,in}$ is averaged between the delayed outlet temperature and the fixed secondary-side temperature $T_{ext,0}$. The reactor parameters correspond to a 3000 MWth MSFR design point \cite{serp_molten_2014}, with nominal fuel temperature $T_F^0 = 973.15$ K, core mass flow rate $\Gamma_0 = 18822$ kg/s, core fuel inventory $M_F = 37129$ kg and prompt neutron generation time $\Lambda = 10^{-6}$ s. The dataset used for MF-SHRED training spans $\tau \in [1, 10]$ s, sampled at 21 equally-spaced values, with each transient integrated up to $T_{final} = 30$ s.

\pagebreak
\bibliographystyle{unsrtsiam}
\bibliography{bibliography}

\end{document}